\documentclass{jfm}

\usepackage[table]{xcolor}

\usepackage{diagbox}

\usepackage{graphicx}
\usepackage{epstopdf, epsfig}
\usepackage{amsmath}
\usepackage[utf8]{inputenc} 

\usepackage{subfigure}

\usepackage{pgfplots}
\usepackage{tikz}
\usetikzlibrary{backgrounds,pgfplots.groupplots,external}

\usepackage{xcolor}

\usepackage{lineno}

\shorttitle{SH instability in nearly--brimful circular--cylinders: a WNL analysis}

\shortauthor{A. Bongarzone, F. Viola, S. Camarri and F. Gallaire}

\title{Sub--harmonic parametric instability\\ in nearly--brimful circular--cylinders:\\ a weakly nonlinear analysis}   

\author{Alessandro Bongarzone\aff{1},
  Francesco Viola\aff{2},
  Simone Camarri\aff{3}
 \and François Gallaire\aff{1}
 \corresp{\email{francois.gallaire@epfl.ch}}}

\affiliation{\aff{1}Laboratory of Fluid Mechanics and Instabilities, École Polytechnique Fédérale de Lausanne, Lausanne, CH-1015, Switzerland
\aff{2}Gran Sasso Science Institute, Viale F. Crispi, 7, 67100 L'Aquila, Italy
\aff{3}Dept. of Industrial and Civil Engineering, Università di Pisa, Pisa, Italy}

\begin{document}

\maketitle

\begin{abstract}

In lab-scale Faraday experiments, meniscus waves respond harmonically  to small-amplitude forcing without threshold, hence potentially cloaking the instability onset of parametric waves. Their suppression can be achieved by resorting to a contact line pinned at the container brim with static contact angle $\theta_s=90^{\circ}$ (brimful condition). However, tunable meniscus waves are desired in some applications as those of liquid-based biosensors, where they can be controlled adjusting the shape of the static meniscus by slightly under/over-filling the vessel ($\theta_s\ne90^{\circ}$) while keeping the contact line fixed at the brim. Here, we refer to this wetting condition as \textit{nearly--brimful}. Although classic inviscid theories based on Floquet analysis have been reformulated for the case of a pinned contact line \citep{Kidambi2013}, accounting for (i) viscous dissipation and (ii) static contact angle effects, including meniscus waves, makes such analyses practically intractable and a comprehensive theoretical framework is still lacking. Aiming at filling this gap, in this work we formalize a  weakly nonlinear analysis via multiple timescale method capable to predict the impact of (i) and (ii) on the instability onset of viscous sub--harmonic standing waves in both brimful and nearly--brimful circular--cylinders. Notwithstanding that the form of the resulting amplitude equation is in fact analogous to that obtained by symmetry arguments \citep{Douady90}, the normal form coefficients are here computed numerically from first principles, thus allowing us to rationalize and systematically quantify the modifications on the Faraday tongues and on the associated bifurcation diagrams induced by the interaction of meniscus and sub--harmonic parametric waves.

\end{abstract}

\begin{keywords}
\end{keywords}


\section{Introduction}\label{sec:Sec1}


When a vessel containing liquid undergoes periodic vertical oscillations, the free liquid surface may be parametrically destabilized with excitation of standing waves depending on the combination of forcing amplitude and frequency. The threshold at which the instability appears is a function of the corresponding mode dissipation and the excited wavelength is generally specified by the wave whose natural frequency is half that of the parametric excitation, as first noticed by \cite{Faraday1831}, who observed experimentally that the resonance was typically of sub--harmonic nature. This observation was later confirmed by \cite{Rayleigh1883a,Rayleigh1883b}, in contrast with \cite{Matthiessen1868,Matthiessen1870}, who observed synchronous vibrations of the free surface with the vertical shaking. The pioneering work of \cite{Benjamin54} gave momentum to the theoretical investigations of the Faraday instability. Using first principles, \cite{Benjamin54} determined that the linear stability of the flat free surface of an ideal fluid within a vertically vibrating container displaying a sliding contact line which intersects orthogonally the container sidewalls is governed by a system of uncoupled Mathieu equations, which predict that standing capillary--gravity waves appear inside the so-called tongues of instability in the driving frequency--amplitude space, with the wave response that can be sub--harmonic, harmonic or super-harmonic, hence reconciling previous observations.\\
\indent The effect of viscous dissipation, taken to be linear and sufficiently small, was initially introduced heuristically \citep{Landau87,Lamb32} in the inviscid solution, resulting in a semi-phenomenological damped Mathieu equation, which was later proved by the viscous linear Floquet theory of \cite{Kumar1994} to be inaccurate, even at small viscosities. An improved version of the damped Mathieu equation, accounting in a more rigorous manner for the dissipation taking place in the free surface and bottom boundary layer, was proposed by \cite{muller1997analytic}, who also noticed in their experiments that the fluid depth can affect the Faraday threshold, with harmonic responses most likely to be triggered for thin fluid layers. The viscous theory of \cite{Kumar1994}, formulated for an horizontally infinite domain, was found to give good agreement with the small-depth large-aspect-ratio experiments of \cite{edwards1994patterns}, where the influence of lateral walls was negligible. If indeed, at large excitation frequencies, where the excited wavelength is much smaller than the container characteristic length, the accessible range of spatial wavenumber is nearly continuous, in the low-frequency regime of single-mode excitation the mode quantization owing to the container sidewall becomes a dominant factor, leading to a discrete spectrum of resonances.\\
\indent A generalization of the viscous Floquet theory to spatially finite systems can be readily obtained by analogy with the inviscid formulation of \cite{Benjamin54}, as \cite{batson2013faraday} recently proposed. It has however intrinsic limitations as it relies on ideal lateral wall conditions, i.e. the unperturbed free surface is assumed to be flat, the contact line is ideally free to slip with a constant zero slope and the stress-free sidewall boundary condition is required for mathematical tractability, since it allows for convenient Bessel-eigenfunctions representation. With the noticeable exception of the sophisticated experiments by \cite{batson2013faraday} and \cite{ward2019faraday} using a gliding liquid coating, these assumptions, by overlooking the contact line dynamics, lead in most experimental cases to a considerable underestimation of the actual overall dissipation, resulting in many cases in an inaccurate prediction of the linear Faraday thresholds in small-container experiments \citep{Benjamin54,dodge1965liquid,ciliberto1985chaotic,henderson1990single,das2008parametrically,tipton2004experimental}. The complexity of the region in the neighborhood of a moving contact line, where molecular, boundary layer and macroscopic scales are intrinsically connected, is indeed of extreme importance and,
despite the significant efforts devoted by several authors to its theoretical understanding \citep{Case1957,Keulegan59,Miles67,Davis1974,Hocking87,miles1990capillary,miles1991capillary,Cocciaro91,Cocciaro93,ting1995boundary,perlin2000capillary,jiang2004contact}, the comparison with moving-contact-line experiments, due to unavoidable sources of uncertainty in the meniscus dynamics, remained mostly qualitative, rather than quantitative, requiring often the use of fitting parameters, e.g. a larger effective fluid viscosity \citep{henderson1990single}.\\
\indent A natural means to get rid of the extra dissipation produced by the contact line dynamics is to simply pin the free surface at the edge of the sidewall, i.e. the container is filled to the brim. In such a condition, the overall dissipation is ruled by that occurring in the fluid bulk and in the Stokes boundary layers at bottom and at the solid lateral walls, where the fluid obeys the classic no-slip boundary condition, relaxing the stress-singularity at the contact line \citep{navier1823memoire,Huh71,Davis1974,miles1990capillary,ting1995boundary,Lauga2007}. Even in the inviscid context, the problem of a pinned contact line boundary condition is well-posed, as shown by the seminal works of \cite{Benjamin79} and \cite{graham1983new}, who first solved the resulting dispersion relation for inviscid capillary--gravity waves with a free surface pinned at the container brim using a variational approach and a suitable Lagrange multiplier. Since then, several semi-analytical techniques, often combining an inviscid solution with boundary layer approximations and asymptotic expansions accounting for viscous dissipation, have been therefore developed to solve the pinned contact line problem, for example in cylindrical containers \citep{henderson1994surface,martel1998surface,miles1998note,nicolas2002viscous,nicolas2005effects,kidambi2009meniscus}. The resulting predictions of natural frequencies and damping coefficients of these capillary--gravity waves, in contradistinction with the case of a moving contact line, showed a remarkable agreement with experimental measurements \citep{henderson1994surface,howell2000measurements}.\\
\indent Within the framework of the Faraday instabilities, this pinned contact line condition can be reached by carefully filling up the vessel up to the brimful condition, as done by \cite{Douady90} and \cite{edwards1994patterns}, among others. Nevertheless, as noticed by \cite{bechhoefer1995experimental}, these delicate experimental conditions are not always perfectly achieved, leading to the presence of a minute meniscus. As mentioned for instance by \cite{Douady90}, the meniscus cannot remain steady upon the oscillating vertical motion of the vessel, which results in the emission of traveling waves from the sidewall to the interior. Irrespective of the pinned or free-edge nature of the contact line, these so-called {\it meniscus waves} are synchronized with the excitation frequency. They are not generated by the parametric resonance, but rather by the modulation of the gravitational acceleration resulting in an oscillating capillary length. They do not need to overcome a minimal threshold in forcing amplitude to appear, are therefore observable in the whole driving frequency-amplitude space and are well described by a purely linear response, i.e. at sufficiently small forcing amplitude,  the meniscus-wave amplitude is proportional to the external forcing amplitude. 
As stated by \cite{Douady90}, edge waves constitute a new time-dependent base state on which the instability of parametric waves may develop, possibly blurring the experimental detection of the true Faraday thresholds. This has led researchers to attempt to suppress edge waves by selecting large-aspect-ratio containers where sidewall effects are negligible, using \textit{sloping sides} or \textit{shelf} conditions to mitigate edge waves by impedance matching \citep{bechhoefer1995experimental}, or employing highly viscous fluid which damp out these waves \citep{Douady90,bechhoefer1995experimental}.\\
\indent With interests in pattern formation, pure meniscus-waves-patterns were investigated for themselves by \cite{torres1995five}, while complex patterns originated by the coupling of meniscus and Faraday waves were recently described by \cite{shao2021role,shao2021surface} for small circular--cylinder experiments. A discussion about harmonic Faraday waves disturbed by harmonic meniscus waves is also outlined in \cite{batson2013faraday}, where the presence of edge waves in a small circular--cylinder--bilayer experiment leads to an imperfect bifurcation diagram, also referred to as a tailing effect by \cite{virnig1988three}, who analyzed sub--harmonic responses only. Interestingly, in some cases, e.g. liquid-based biosensors for DNA detection \citep{picard2007resonance}, tunable small-amplitude stationary waves as meniscus waves are actually desired and preferred to saturated larger-amplitude Faraday waves. In such applications, a starting brimful condition, having a contact line fixed at the brim, is ideal since the effective static contact angle at the wall and hence the size and shape of the static meniscus, which will emit edge waves under vertical excitations, can be adjusted simply by increasing or decreasing the bulk volume (\textit{nearly--brimful} condition).\\
\indent Although the non-conventional eigenvalue problem for natural frequencies and damping coefficients of pinned-contact-line capillary--gravity waves was tackled by several authors mentioned above and in spite of the vastness of literature focused on Faraday waves, there is lack of a comprehensive theoretical framework for the investigation of such a configuration within the context of Faraday instability. An important exception is the work of \cite{Kidambi2013}. Assuming inviscid Faraday waves in a brimful--cylinder with a ideally flat static free surface, he represented the problem using appropriate modal solutions followed by a projection on a test function space and showed that pinned contact line condition resulted in an infinite system of coupled Mathieu equations, unlike the classic case of an ideal moving contact line \citep{Benjamin54}. Nevertheless, viscosity, crucial for an accurate prediction of the Faraday threshold, was not included in the analysis, nor was the presence of a static meniscus and its consequent emission of meniscus waves. Some attempts to include meniscus modifications to the Faraday thresholds have been made by several authors by including periodic inhomogeneities \citep{ito1999interface,tipton2003interfacial} and \textit{ad hoc} phenomenological terms \citep{lam2011effect} to an \textit{ad hoc} damped Mathieu equation.\\
\indent The purpose of this work is to take one more step in the direction undertaken by \cite{Kidambi2013}, by rigorously accounting for (i) viscous damping, (ii) a pinned contact line and (iii) the presence of a static meniscus at rest. As mentioned above, a contact-angle different from 90 degrees not only results in a static meniscus, but also induces the emission of meniscus waves as the static meniscus shape is no longer a solution of the forced problem, even below Faraday threshold. A Floquet-inspired linear theory \textit{à la} \cite{Kumar1994} cannot be pursued, as perturbations develop around an oscillating base-flow. In contrast, we propose to use the weakly nonlinear approach (WNL) to approximate the linear Faraday bifurcations, although it is expected to involve cumbersome calculations.\\
\indent Weakly nonlinear analyses \citep{Miles84,Meron86,Nayfeh1987,nagata1989nonlinear,henderson1990single,Milner91,Douady90,Zhang97,chen1999amplitude,skeldon2007pattern,jian2005instability,Rajchenbach2015} have indeed been widely used in the context of Faraday instabilities to study the wave amplitude saturation via super- and subcritical bifurcations, as well as to investigate pattern and quasi-pattern formation \citep{edwards1993parametrically,edwards1994patterns} or  spatiotemporal chaos \citep{ciliberto1985chaotic,gluckman1993time}, arising when two modes with nearly the same frequency share the same unstable region in the parameter space and strongly interact. In contradistinction with these previous studies, the presence of a static meniscus calls for a WNL approach not only to estimate the wave amplitude saturation in the weakly nonlinear regime, but also to predict the Faraday threshold. Hence, with regard to cylindrical straight-sidewalls and sharp-edged containers, as the one considered by \cite{shao2021surface}, we derive a WNL model capable to simultaneously account for viscous dissipation, static meniscus and meniscus waves, thus allowing us to predict their influence on the linear Faraday threshold for standing capillary--gravity waves with  pinned contact line as well as their saturation to finite amplitude. Following the recent experimental evidences of \cite{shao2021surface}, we focus on single-mode sub--harmonic resonances. To this end, the full system of equations governing the fluid motion is solved asymptotically by means of the method of multiple timescales, involving a series of linear problems, which are solved numerically.  The theoretical model results in a final amplitude equation for the wave amplitude, $B$, whose form corresponds to that derived by \cite{Douady90} using symmetry arguments solely and keeping low order terms only,
\begin{equation}
\label{eq:amp_eq_intro}
\frac{dB}{d t}=-\left(\sigma+\text{i}\Lambda/2\right)B+\zeta FB^*+\chi|B|^2B.
\end{equation}
\noindent The form of~\eqref{eq:amp_eq_intro} is indeed valid whatever the shape of the static surface, mode structure and the boundary condition are, but the normal form coefficients, which account for the effect of the static contact angle and which are complex values owing to the presence of viscosity, are here formally determined in closed form from first principles and computed numerically.\\
\indent The paper is organized as follows. In \S\ref{sec:Sec2} the flow configuration and governing equations are introduced, while the numerical methods and tools employed in the work are presented in \S\ref{sec:Sec3}. In \S\ref{sec:Sec4} we formulate a linear eigenvalue problem for the damping and natural frequency of viscous capillary--gravity waves with pinned contact line, whose numerical solution is then compared with several previous experiments and theories. The WNL model for sub--harmonic Faraday resonances is formalized in \S\ref{sec:Sec5}. A \textit{vis}-\textit{à}-\text{vis} comparison with recent experiments by \cite{shao2021surface} with a pure brimful configuration are discussed before moving to a systematic investigation of meniscus effects. Lastly, for validation purposes, in \S\ref{sec:Sec6} the modified bifurcation diagram presented in \S\ref{sec:Sec5} is compared for a specific case, i.e. pure axisymmetric dynamics, with fully nonlinear direct numerical simulation (DNS). Final comments and conclusions are outlined in \S\ref{sec:Sec7}.


\section{Flow Configuration and governing equations}\label{sec:Sec2}

\begin{figure}
\centering
\subfigure{%
  \includegraphics[scale=0.275]{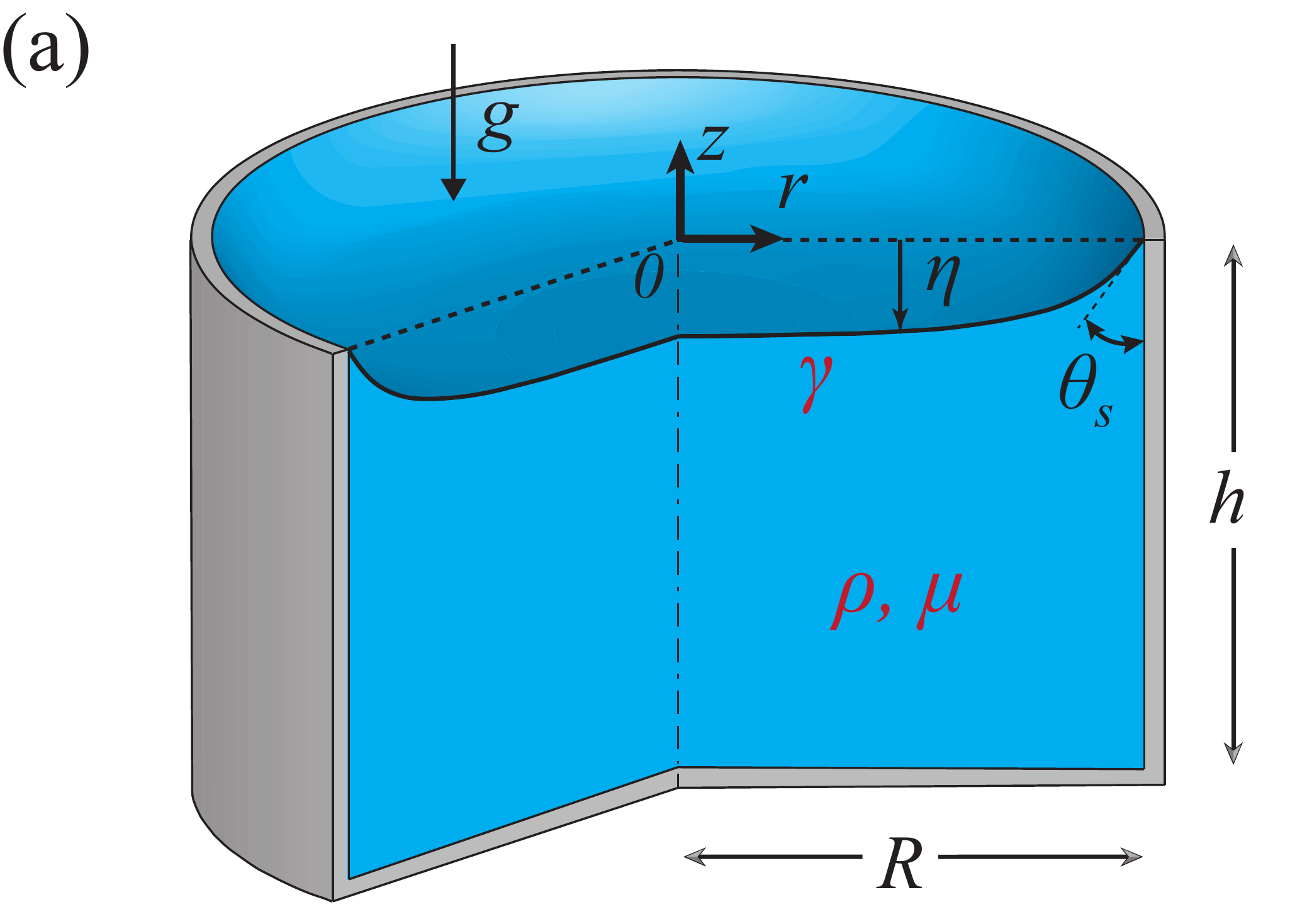}
  \label{fig:subfigure1}}
\quad
\subfigure{%
  \includegraphics[scale=0.275]{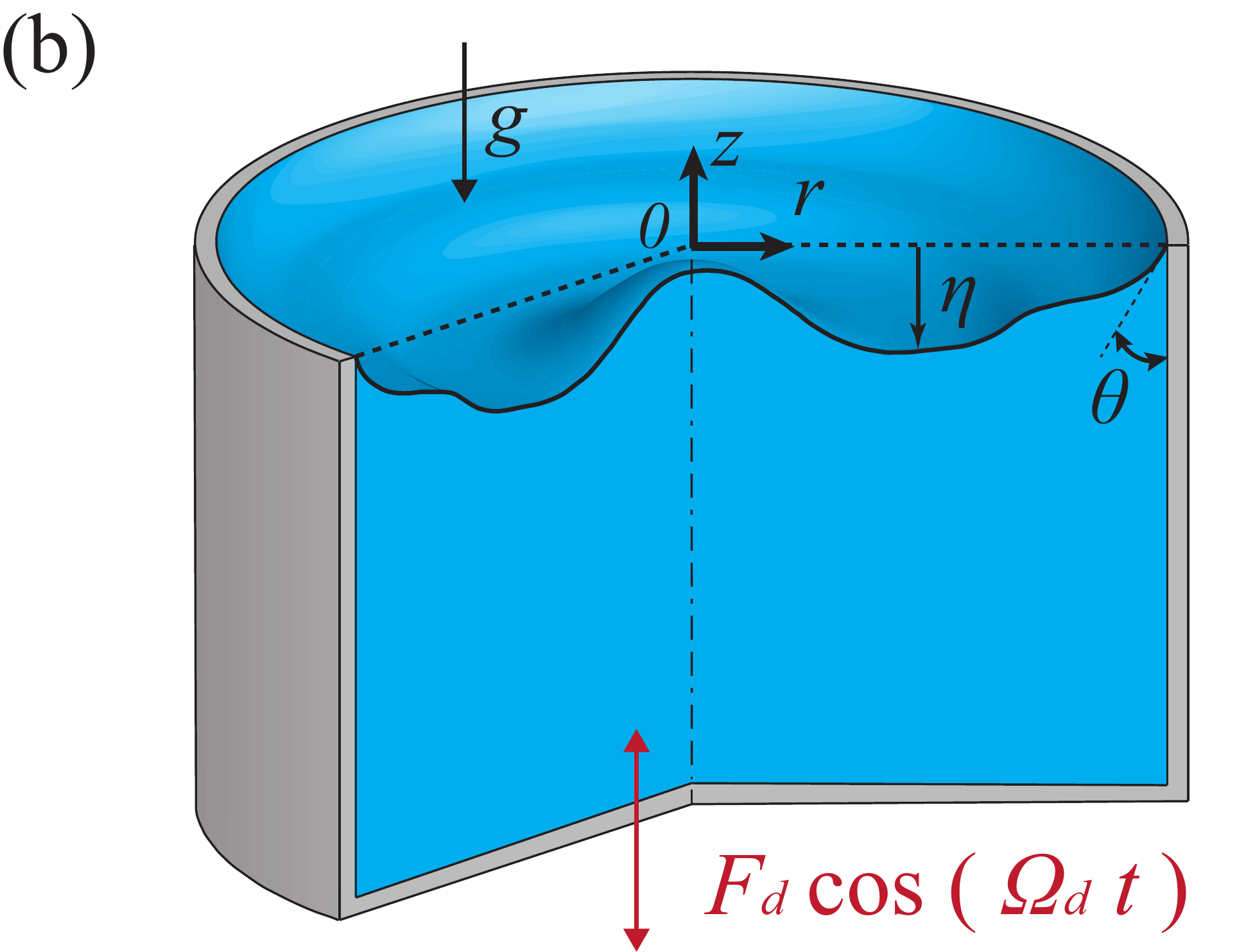}
  \label{fig:subfigure2}}
\caption{Sketch of a straight-sidewalls sharp-edged cylindrical container of radius $R$ and filled to a depth $h$ with a liquid of density $\rho$ and
dynamic viscosity $\mu$. The air-liquid surface tension is denoted by $\gamma$. (a) The free surface, $\eta$, is represented in a generic static configuration characterized by a static contact angle $\theta_s$. (b) Generic dynamic configuration under the external vertical periodic forcing of amplitude $F_d$ and angular frequency $\Omega_d$. The contact line is pinned. $rz$-plane: reference working plane.}
\label{fig:sketch_geom}
\end{figure}

We consider a cylindrical vessel of radius $R$ and filled to a depth $h$ with a liquid of density $\rho$ and
dynamic viscosity $\mu$ (see figure~\ref{fig:sketch_geom}). The vessel undergoes a vertical periodic acceleration $F_d=A_d\Omega_d^2$, where $A_d$ and $\Omega_d=2\pi f_d$ are the driving amplitude and angular frequency, respectively. In a non--inertial reference frame, the fluid experiences a vertical acceleration due to the unsteady apparent gravitational acceleration $g_{app}\left(t\right)=g\left[1-\left(F_d/g\right)\cos{\Omega_d}\,t\right]$. The viscous fluid motion is thus governed by the incompressible Navier--Stokes equations, 
\begin{equation}
\label{eq:GovEqNS}
\nabla\cdot\mathbf{u}=0\ \ \ \ \ ,\frac{\partial \mathbf{u}}{\partial t}+\left(\mathbf{u}\cdot\nabla\right)\mathbf{u}+\nabla p-\frac{1}{Re}\Delta\mathbf{u}=-\left(1-\frac{F_d}{g}\cos{\Omega_d\,t}\right)\hat{\mathbf{e}}_z,
\end{equation}

\noindent with $\mathbf{u}=\left\{u_r,u_{\phi},u_z\right\}^T$ velocity field and $p$ pressure field. Equations~\eqref{eq:GovEqNS} are made nondimensional by using the container's characteristic length $R$, the characteristic velocity $\sqrt{gR}$ and the time scale $\sqrt{R/g}$. The pressure gauge is set to $\rho g R$. Consequently, the Reynolds number is defined as $Re=\rho g^{1/2} R^{3/2}/\mu$ and the term on the r.h.s. represents the time--modulation of the nondimensional gravity acceleration. At the interface $z=\eta$ we impose the kinematic and dynamic boundary conditions (b.c.), 

\begin{subequations}
\label{eq:bcKinDyn}
\begin{equation}
\label{eq:bcKin}
\frac{\partial\eta}{\partial t}+\left.\mathbf{u}\right|_{\eta}\cdot\mathbf{n}\left(\eta\right)=0,
\end{equation}

\begin{equation}
\label{eq:bcDyn}
-\left.p\right|_{\eta}\mathbf{n}\left(\eta\right)+\frac{1}{Re}\left.\left(\nabla\mathbf{u}+\nabla^T\mathbf{u}\right)\right|_{\eta}\cdot\mathbf{n}\left(\eta\right)=\frac{1}{Bo}\kappa\left(\eta\right)\mathbf{n}\left(\eta\right),
\end{equation}
\end{subequations}

\medskip
\noindent where $\eta$ denotes the interface coordinate, $\kappa\left(\eta\right)$ is the free surface curvature, $\mathbf{n}\left(\eta\right)$ is unit vector locally normal to the interface and $Bo$ is the Bond number defined as $Bo=\rho gR^2/\gamma$, with $\gamma$ air--liquid surface tension. At the solid bottom, $z=-h/R=-H$ and sidewall, $r=1$, we impose the no--slip b.c., $\mathbf{u}=\mathbf{0}$. Lastly, the dynamic pinned (or fixed) contact line condition is enforced as 

\begin{equation}
\label{eq:PinnBC}
\left.\frac{\partial\eta}{\partial t}\right|_{r=1}=0.
\end{equation}


\section{Numerical methods and tools}\label{sec:Sec3}

Different numerical approaches are adopted in the present paper. The numerical scheme used in the eigenvalue calculation, \S\ref{sec:Sec4}, and in the weakly nonlinear analysis, \S\ref{sec:Sec5}, is a staggered Chebyshev--Chebyshev collocation method implemented in Matlab. The three velocity components are discretized using a Gauss--Lobatto--Chebyshev (GLC) grid, whereas the pressure is staggered on Gauss--Chebyshev (GC) grid. Accordingly, the momentum equation is collocated at the GLC nodes and the pressure is interpolated from the GC to the GLC grid, while the continuity equation is collocated at the GC nodes and the velocity components are interpolated from the GLC to the GC grid. This results in the classical $P_N$-$P_{N-2}$ formulation, which automatically suppress spurious pressure modes in the discretized problem. A two-dimensional mapping is then used to map the computational space onto the physical space, that has, in general, a curved boundary due to the presence of concave or convex static meniscus. Lastly, the partial derivatives in the computational space are mapped onto the derivatives in the physical space, which depend on the mapping function. For other details see \cite{heinrichs2004spectral,canuto2007spectral,sommariva2013fast,Viola2016a,Viola2018a,Viola2018b}.\\
\indent The weakly nonlinear model presented in \S\ref{sec:Sec5} involves a third order asymptotic expansion of the full hydrodynamic system introduced in \S\ref{sec:Sec2}, that turns out to be very tedious to derive analytically. Therefore, the linearization and expansion procedures have been fully automated using the software Wolfram Mathematica, a powerful tool for symbolic calculus, which has been then integrated within the main Matlab code. The Mathematica codes are provided in the supplementary material as a support to the readers.\\
\indent In \S\ref{sec:Sec6}, the results obtained from the weakly nonlinear analysis are compared and validated for a specific case, i.e. axisymmetric dynamics, with axisymmetric and fully nonlinear direct numerical simulations (DNS), which have been performed using the finite-element software COMSOL Multiphysics v5.6. Further details about the specific DNS setting will be given in \S\ref{sec:Sec6}.


\section{Damping and frequency of capillary--gravity waves with pinned contact line: comparison with previous analyses and experiments}\label{sec:Sec4}

Assuming at first the case with zero external forcing, in this section we study the damping and natural frequencies of viscous capillary--gravity waves with fixed contact line and we compare our numerical results with existing experiments and with previous theoretical and numerical predictions. To this end, the flow field $\mathbf{q}=\left\{\mathbf{u},p\right\}^T$ and the interface $\eta$ are decomposed in a static base flow, $\mathbf{q}_0=\left\{\mathbf{0},p_0\right\}^T$ and $\eta_0$, and a small perturbation, $\mathbf{q}_1=\left\{\mathbf{u}_1,p_1\right\}^T$ and $\eta_1$, of infinitesimal amplitude $\epsilon$, i.e. $\mathbf{q}=\mathbf{q}_0+\epsilon\mathbf{q}_1$ and $\eta=\eta_0+\epsilon\eta_1$.

\subsection{Static meniscus}\label{subsec:Sec4subsec1}

At rest, the velocity field $\mathbf{u}_0$ is null everywhere and the pressure is hydrostatic, i.e. $p_0=-z$. Therefore, the static configuration is obtained by solving the nonlinear equation associated with the shape of the axisymmetric static meniscus, $\eta_0\left(r\right)$,

\begin{equation}
\label{eq:static_men}
\eta_0-\frac{\kappa\left(\eta_0\right)}{Bo}=0.
\end{equation}

\noindent with $\kappa\left(\eta_0\right)=\left(\eta_{0,rr}+\eta_{0,r}\left(1+\eta_{0,r}^2\right)/r\right)\left(1+\eta_{0,r}^2\right)^{-3/2}$. At the centerline, $r=0$, the regularity condition $\eta_{0,r}=0$ holds owing to axisymmetry. The shape of the meniscus is obtained by imposing the geometric relation at the contact line, $r=1$,

\begin{equation}
\label{eq:static_men_geom_rel}
\left.\frac{\partial\eta_0}{\partial r}\right|_{r=1}=\cot{\theta_s},
\end{equation}

\noindent where $\theta_s$ is a prescribed static contact angle (see also figure~\ref{fig:sketch_geom}(a)). When $\theta_s$ is set to $\pi/2$, then the static interface appears flat.

\subsection{Linear eigenvalue problem}\label{subsec:Sec4subsec2}

Governing equations~\eqref{eq:GovEqNS} and their boundary conditions~\eqref{eq:bcKinDyn} are then linearized around the static base-flow. It follows that at order $\epsilon$ the velocity and pressure fields satisfy the Stokes equations
\begin{equation}
\label{eq:GovEqNS_eps1}
\nabla\cdot\mathbf{u}_1=0,\ \ \ \ \ \frac{\partial \mathbf{u}_1}{\partial t}+\nabla p_1-\frac{1}{Re}\Delta\mathbf{u}_1=\mathbf{0},
\end{equation}

\noindent with the linearized kinematic and dynamic free surface boundary conditions (at $z=\eta_0$)

\begin{equation}
\label{eq:bcKin_eps1}
\frac{\partial\eta_1}{\partial t}+\left.\mathbf{u}_1\right|_{\eta_0}\cdot\mathbf{n}\left(\eta_0\right)=0,
\end{equation}
\begin{equation}
\label{eq:bcDyn_eps1}
-\left.p_1\right|_{\eta_0}\mathbf{n}\left(\eta_0\right)+\eta_1\mathbf{n}\left(\eta_0\right)+\frac{1}{Re}\left.\left(\nabla\mathbf{u}_1+\nabla^T\mathbf{u}_1\right)\right|_{\eta_0}\cdot\mathbf{n}\left(\eta_0\right)=\frac{1}{Bo}\left.\frac{\partial\kappa\left(\eta\right)}{\partial\eta}\right|_{\eta_0}\eta_1\mathbf{n}\left(\eta_0\right),
\end{equation}

\noindent where $\mathbf{n}\left(\eta_0\right)=\left\{-\eta_{0,r},0,1\right\}^T\left(1+\eta_{0,r}^2\right)^{-1/2}$ and 

\begin{equation}
\label{eq:Curv_eps1}
\left.\frac{\partial\kappa\left(\eta\right)}{\partial\eta}\right|_{\eta_0}\eta_1 = \frac{\left(1+\eta_{0,r}^2\right)-3r\eta_{0,r}\eta_{0,rr}}{\left(1+\eta_{0,r}^2\right)^{5/2}}\frac{1}{r}\frac{\partial\eta_1}{\partial r}+\frac{1}{\left(1+\eta_{0,r}^2\right)^{3/2}}\frac{\partial^2\eta_1}{\partial r^2}+\frac{1}{\left(1+\eta_{0,r}^2\right)^{1/2}}\frac{1}{r^2}\frac{\partial^2\eta_1}{\partial\phi^2}
\end{equation}

\noindent is the first order variation of the curvature associated with the small perturbation $\epsilon\eta_1$. The azimuthal coordinated is denoted by $\phi$. The no-slip boundary condition is imposed at the solid walls, $\mathbf{u}_1=\mathbf{0}$, while the pinned contact line condition is enforced at the contact line, $z=\eta_0$ and $r=1$,

\begin{equation}
\label{eq:BcCLpinn}
\left.\frac{\partial\eta_1}{\partial t}\right|_{r=1}=0.
\end{equation}

\noindent Hence, the linear system can be written in compact form as

\begin{equation}
\label{eq:eps1_AB}
\left(\mathcal{B}\partial_t-\mathcal{A}\right)\mathbf{q}_1=\mathbf{0},\ \ \ \text{with}\ \ \ 
\mathcal{B}=\left(
\begin{matrix}
I & 0\\
0 & 0
\end{matrix}
\right),\ \ \ 
\mathcal{A}=\left(
\begin{matrix}
Re^{-1}\Delta & -\nabla\\
\nabla^T & 0
\end{matrix}
\right).
\end{equation}

\begin{table}
\centering
\begin{tabular}{c|cc|cc}
Literature survey  & meniscus-free ($\theta_s=90^{\circ}$) & Acr. &  with meniscus ($\theta_s\ne90^{\circ}$) & Acr.\\ 
& & &\\ \hline
Experimental & \cite{henderson1994surface} & HM94 & \cite{Cocciaro93} & C93\\ 
campaigns & \cite{howell2000measurements} & H2000 & \cite{picard2007resonance} & PD07\\ \hline
 & \cellcolor{gray!15}\cite{henderson1994surface} & \cellcolor{gray!15}HM94 & \cellcolor{gray!15} & \cellcolor{gray!15}\\
Viscous & \cellcolor{gray!15}\cite{martel1998surface} & \cellcolor{gray!15}M98 & \cellcolor{gray!15} & \cellcolor{gray!15} \\ 
analyses & \cellcolor{gray!15}\cite{miles1998note} & \cellcolor{gray!15}MH98 & \cellcolor{gray!15}\cite{kidambi2009meniscus} & \cellcolor{gray!15}K09 \\ 
 & \cellcolor{gray!15}\cite{nicolas2002viscous} & \cellcolor{gray!15}N02  & \cellcolor{gray!15} & \cellcolor{gray!15} \\ 
 & \cellcolor{gray!15}\cite{kidambi2009meniscus} & \cellcolor{gray!15}K09 & \cellcolor{gray!15} & \cellcolor{gray!15} \\ \hline
  & \cite{graham1983new} & GE83 &  & \\
Inviscid  & \cite{henderson1994surface} & HM94 &  & \\
 analyses &\cite{Kidambi2013}  & K13 & \cite{nicolas2005effects} &  N05\\
  & \cite{shao2021surface} & S21 &  &  \\
\end{tabular}
\caption{Literature survey on the natural frequencies and damping coefficients of small-amplitude capillary--gravity waves in lab-scale upright cylindrical containers with pinned contact line and in both meniscus-free and with-meniscus configurations. The present work lies within the conditions highlighted by the shaded frames. The case examined by K13 and S21 will be discussed afterwards in \S\ref{sec:Sec5} within the context of sub--harmonic Faraday waves.}
\label{tab:literature_tab}
\end{table}

\noindent We note that the kinematic and the dynamic b.c.s~\eqref{eq:bcKin_eps1} and~\eqref{eq:bcDyn_eps1} do not explicitly appear in~\eqref{eq:eps1_AB}, but they are enforced as conditions at the interface \citep{Viola2018b}. In practice, in the numerical scheme an additional variable, $\eta$, is added to~\eqref{eq:eps1_AB}. The free surface $\eta$ is therefore dynamically coupled with $\mathbf{u}_1$ and $p_1$ and the solution can be expanded in terms of normal modes in time and in the azimuthal direction
\begin{equation}
\label{eq:norm_mode_eig}
\mathbf{q}_1\left(r,\phi,z,t\right)=\hat{\mathbf{q}}_1\left(r,z\right)e^{\lambda t}e^{\text{i}m\phi}+c.c., \ \ \ \ \ \ \eta_1\left(r,\phi,t\right)=\hat{\eta}_1\left(r\right)e^{\lambda t}e^{\text{i}m\phi}+c.c.
\end{equation}

\noindent Substituting the normal form~\eqref{eq:norm_mode_eig} in system~\eqref{eq:eps1_AB}, we obtain a generalized linear eigenvalue problem,
\begin{equation}
\label{eq:eig_eps1}
\left(\lambda\mathcal{B}-\mathcal{A}_m\right)\hat{\mathbf{q}}_1=\mathbf{0},
\end{equation}

\noindent where the linear operator $\mathcal{A}_m$ depend on the azimuthal wavenumber $m$ and $\hat{\mathbf{q}}_1$ is the global mode associated with the eigenvalue $\lambda=-\sigma+\text{i}\omega$, with $\sigma$ and $\omega$ the damping coefficient and the natural frequency, respectively, of the $\left(m,n\right)$ global mode. Here the indices $\left(m,n\right)$ represent the number of nodal circles and nodal diameters, respectively. Owing to the normal mode expansion~\eqref{eq:norm_mode_eig}, we notice that the operator $\mathcal{A}_m$ is complex, since $\phi$ derivatives produce $\text{i}m$ terms. A complete expansion of the complex operator can be found in \cite{Meliga2009} and \cite{Viola2018b}.\\
\indent In order to regularize the problem at the axis, depending on the selected azimuthal wavenumber $m$, different regularity conditions must be imposed at $r=0$ \citep{liu2012nonmodal,Viola2018b},
\begin{subequations}
\begin{equation}
\label{eq:regularity_cond_m0}
m=0\text{:}\ \ \ \ \ \ \hat{u}_{1r}=\hat{u}_{1\phi}=\frac{\partial\hat{u}_{1z}}{\partial r}=\frac{\partial \hat{p}_1}{\partial r}=0,
\end{equation}
\begin{equation}
\label{eq:regularity_cond_m1}
|m|=1\text{:}\ \ \ \ \  \frac{\partial\hat{u}_{1r}}{\partial r}=\frac{\partial\hat{u}_{1\phi}}{\partial r}=\hat{u}_{1z}=\hat{p}_1=0,
\end{equation}
\begin{equation}
\label{eq:regularity_cond_m234}
|m|>0\text{:}\ \ \ \ \ \ \ \ \ \ \hat{u}_{1r}=\hat{u}_{1\phi}=\hat{u}_{1z}=\hat{p}_1=0.
\end{equation}
\end{subequations}

\noindent Lastly, we underly that owing to the symmetries of the problem, system~\eqref{eq:eig_eps1} is invariant under the 
\begin{equation}
\label{eq:m_transf}
\left(\hat{u}_{1r},\hat{u}_{1\phi},\hat{u}_{1z},\hat{p}_1,\hat{\eta}_1,+m,-\sigma+\text{i}\omega\right)\rightarrow\left(\hat{u}_{1r},-\hat{u}_{1\phi},\hat{u}_{1z},\hat{p}_1,\hat{\eta}_1,-m,-\sigma+\text{i}\omega\right),
\end{equation}

\medskip
\noindent transformation, so in this section, \S\ref{sec:Sec4}, we consider only the case with $m\geqslant 0$. Furthermore, while $\left(\hat{\mathbf{q}}_1,\hat{\eta}_1,m,-\sigma+\text{i}\omega\right)$ is solution of~\eqref{eq:eig_eps1}, $\left(\hat{\mathbf{q}}_1^*,\hat{\eta}_1^*,m,-\sigma-\text{i}\omega\right)$ (where the star designates the complex conjugate) is not a solution, instead the following relation holds

\begin{equation}
\label{eq:omega_transf1}
\left(\hat{\mathbf{q}}_1,\hat{\eta}_1,+m,-\sigma+\text{i}\omega\right)\rightarrow\left(\hat{\mathbf{q}}_1^*,\hat{\eta}_1^*,-m,-\sigma-\text{i}\omega\right),
\end{equation}
\begin{equation}
\label{eq:omega_transf2}
\left(\hat{\mathbf{q}}_1,\hat{\eta}_1,-m,-\sigma+\text{i}\omega\right)\rightarrow\left(\hat{\mathbf{q}}_1^*,\hat{\eta}_1^*,+m,-\sigma-\text{i}\omega\right),
\end{equation}

\noindent i.e. the eigenvalues are complex conjugates and all spectra ($\pm m$) in the $\left(\sigma,\omega\right)$--plane are symmetric with respect to the real axis ($\omega=0$), but the complex conjugates of the corresponding eigenvectors, with the exception for axisymmetric dynamics ($m=0$), are not eigenmodes of the same spectrum. The damping coefficients and natural frequencies of viscous capillary--gravity waves with fixed contact line in both the meniscus-free and with-meniscus configuration are thus computed by solving numerically the generalized eigenvalue problem~\eqref{eq:eig_eps1}, as described in \S\ref{sec:Sec3}.\\
\indent With regard to the literature survey outlined in table~\ref{tab:literature_tab}, in the following, we propose a thorough validation of our numerical tools via comparison with several pre-existing experiments and theoretical/semi-analytical predictions.

\subsection{Flat static free surface: $\theta_s=90^{\circ}$}\label{subsec:Sec4subsec3}

Let us start by considering the case of a flat static interface, i.e. the static contact angle is set to $\theta_s=90^{\circ}$, for which $\eta_0\left(r\right)=0$ (perfect brimful condition).

\subsubsection{Experiments and theories by HM94, MH98 and M98}\label{subsubsec:Sec4subsec3subsubsec1}

\begin{figure}
    \centering
    \includegraphics[width=1.\textwidth]{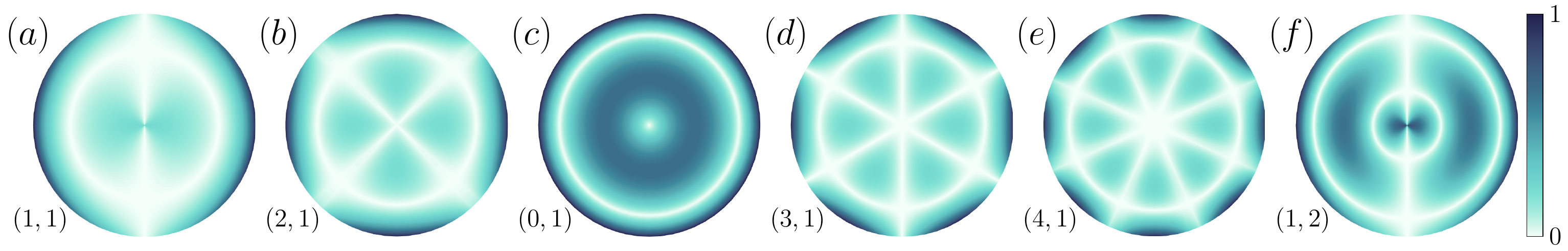}
    \caption{Shape of the eigensurfaces associated with the six global modes considered in table~\ref{tab:expHM} and denoted by the indices $\left(m,n\right)$. The magnitude of the eigensurface slope is plotted. The eigenmodes are normalized with the phase and absolute value of the slope at the contact line.}
    \label{fig:six_eigenfree_HM}
\end{figure}

We consider here the experimental measurements by HM94 for the first six modes in a brimful, sharp-edged cylinder in absence of free surface contamination. The corresponding geometrical and fluid properties are reported in caption of table~\ref{tab:expHM}, while the eigensurfaces associated with the first six modes, computed by solving numerically the eigenvalue problem~\eqref{eq:eig_eps1}, are shown in figure~\ref{fig:six_eigenfree_HM}.\\
\indent In table~\ref{tab:expHM}, the experimental damping coefficients and angular frequencies measured by HM94 are compared with their own viscous theoretical predictions, with the prediction of M98 for the very same case and with our numerical results. If the frequency prediction of HM94 is in good agreement with their own experiments, a significant mismatch is found in terms of damping coefficient. However, this discrepancy is strongly reduced in the prediction of M98, which is in agreement with our numerical results. By analogy with M98, the theory proposed in HM94 was supplemented in MH98 by a calculation of the interior damping (based on Lamb’s dissipation integral for an irrotational flow \citep{Lamb32}), which yields results (here omitted for the sake of brevity) of comparable accuracy with M98 and with the present predictions. We note that the predicted frequencies in both M98 and the present study are always within 0.3\% of the experimental values.

\begin{table}
\centering
\begin{tabular}{c|cc|ccc|ccc|ccc}
& \multicolumn{2}{c|}{Exp. HM94} & \multicolumn{3}{c|}{Theory HM94} & \multicolumn{3}{c|}{Theory M98} & \multicolumn{3}{c}{Present Num.}\\
$\left(m,n\right)$ & $f_E$ $\left(\text{Hz}\right)$ & $\Delta_E$ $\left(-\right)$ & $f_T$ & $\Delta_T$ & $\Delta_E/\Delta_T$ & $f_T$ & $\Delta_T$ & $\Delta_E/\Delta_T$ & $f$ & $\Delta$ & $\Delta_E/\Delta$\\ \hline
$\left(1,1\right)$ & 4.65 & $ 1.4$ & 4.66 & 1.13 & 1.2 & 4.67 & 1.37 & 1.02 & 4.66 & 1.36 & 1.03\\
$\left(2,1\right)$ & 6.32 & $ 1.8$ & 6.32 & 1.24 & 1.4 & 6.34 & 1.75 & 1.03 & 6.34 & 1.74 & 1.03\\
$\left(0,1\right)$ & 6.84 & $ 1.2$ & 6.73 & 0.44 & 2.7 & 6.85 & 0.95 & 1.26 & 6.85 & 0.93 & 1.29\\
$\left(3,1\right)$ & 7.80 & $ 2.2$ & 7.79 & 1.29 & 1.7 & 7.82 & 2.11 & 1.04 & 7.82 & 2.08 & 1.06\\
$\left(4,1\right)$ & 9.26 & $ 2.4$ & 9.24 & 1.32 & 1.8 & 9.27 & 2.47 & 0.97 & 9.27 & 2.42 & 0.99\\
$\left(1,2\right)$ & 8.57 & $ 1.5$ & 8.57 & 0.48 & 3.1 & 8.59 & 1.45 & 1.03 & 8.59 & 1.43 & 1.05\\
\end{tabular}
\caption{Experimental frequency and damping by HM94, their theoretical prediction and the theoretical prediction by M98 are compared with the present numerical results. Geometrical and fluid properties: $R=0.02766\,\text{m}$, $h=0.038\,\text{m}$, $\rho=1000\,\text{kgm}^{-3}$, $\mu=0.001\,\text{kgm}^{-1}\text{s}^{-1}$, $\gamma=0.0724\,\text{Nm}^{-1}$, for which $Re=14\,401$ and $Bo=103.6$, and a static angle $\theta_s=90^{\circ}$. The dimensionless damping coefficient $\sigma$ is rescaled according to HM94, i.e. $\Delta=4\sqrt{Re/2\omega}\sigma$, where $\sigma$ and $\omega$ for the present numerical results (last three columns) are those computed by solving~\eqref{eq:eig_eps1}. The dimensional frequency is readily obtained as $f=\left(\omega/2\pi\right)\sqrt{g/R}$. The number of points in the radial and axial directions for the GLC grid used is this calculation is $N_r=N_z=40$, for which convergence is achieved.}
\label{tab:expHM}
\end{table}

\subsubsection{Experiments and theories by H2000, M98, N02 and K09}\label{subsubsec:Sec4subsec3subsubsec2}

Table~\ref{tab:expH} provides a comparison of the present results with the experimental measurements of H2000, the asymptotic calculations of M98, the theoretical predictions of N02 and the calculations of K09.\\
\indent All the theoretical methods accurately predict the natural frequencies, even at low $Re$, as the viscous correction is very small. However, in terms of damping, it is seen that the asymptotic model of M98 is increasingly inaccurate for decreasing $Re$. For the present case, our numerical calculations place in between N02 and K09, with frequency predictions within 0.7\% of the experimental values.

\begin{table}
\centering
\begin{tabular}{c|cc|cc|cc|cc|cc}
& \multicolumn{2}{c|}{Exp. H2000} & \multicolumn{2}{c|}{Theory M98} & \multicolumn{2}{c|}{Theory N02} & \multicolumn{2}{c|}{Num. K09} & \multicolumn{2}{c}{Present Num.}\\
$Re$ & $f_E$ $\left(-\right)$ & $\Delta_E$ $\left(-\right)$ & $f_T/f_E$ & $\Delta_T/\Delta_E$ & $f_T/f_E$ & $\Delta_T/\Delta_E$ & $f_N/f_E$ & $\Delta_N/\Delta_E$ & $f/f_E$ & $\Delta/\Delta_E$\\ \hline
13\,077.02 & 2.079 & 0.0052 & 1.004 & 0.984 & 1.005 & 0.911 & 1.005 & 0.920 & 1.005 & 0.911\\
6\,422.61 & 2.075 & 0.0088 & 1.005 & 0.984 & 1.007 & 0.942 & 1.007 &0.954 & 1.007 & 0.947\\
2\,620.55 & 2.075 & 0.0181 & 1.005 & 1.040 & 1.006 & 0.968 & 1.006 & 0.971 & 1.006 & 0.967\\
1\,317.35 & 2.072 & 0.0332 & 1.006 & 1.046 & 1.006 & 0.945 & 1.006 & 0.949 & 1.006 & 0.948\\
575.37 & 2.066 & 0.0660 & 1.008 & 1.135 & 1.005 & 0.975 & 1.006 & 0.979 & 1.005 & 0.978\\
269.91 & 2.059 &0.1271 & 1.010 & 1.193 & 1.001 & 0.979 & 1.001 & 0.982 & 1.001 & 0.981\\
\end{tabular}
\caption{Dimensionless damping and frequency of the first axisymmetric mode $\left(0,1\right)$ for different $Re$. Nondimensional parameters: $R=1$, $h/R=1.379$, $Bo=365$ and $\theta_s=90^{\circ}$. Here the dimensionless natural frequency and damping correspond to $f=\omega$ and $\Delta=\sigma$ in our notation. The number of points in the radial and axial directions for the GLC grid used is this calculation is $N_r=N_z=40$, for which convergence is achieved. Comparisons outlined in this table (except for last column) are provided in table~2 of K09.}
\label{tab:expH}
\end{table}

\subsection{Presence of static meniscus: $\theta_s \ne 90^{\circ}$}\label{subsec:Sec4subsec4}

We now analyze the case of an initially non-flat static interface, i.e. $\theta_s\ne90^{\circ}$, for which $\eta_0\left(r\right)\ne0$ (nearly--brimful condition), and its effect on the natural frequencies and damping coefficients of viscous capillary--gravity waves with a pinned contact line. 

\subsubsection{Experiments by C93 and calculations by N05 and K09}\label{subsubsec:Sec4subsec4subsubsec1}

C93 measured the frequency and damping rate of the first non-axisymmetric mode $\left(m,n\right)=\left(1,1\right)$ in a cylindrical container where the static free surface had an effective static contact angle $\theta_s=62^{\circ}$. They identified two different regimes, namely, a higher and a smaller amplitude regime. In the latter, the contact line was observed to remain pinned. N05 and K09 have computed the damping and frequency for this case and a comparison with our numerical analysis is reported in table~\ref{tab:expC}. We note that the prediction of N05 is close to the experimental values, however such a prediction is based on an asymptotic representation of the static meniscus, while in the present calculation, as well as the one proposed by K09, it is computed numerically. Moreover, the damping prediction by N05 relies on HM94 and M98 theories, since his starting point is an inviscid analysis. Our result seems to be slightly closer to the experimental values than the one of K09, although both are in fairly good agreement.

\begin{table}
\centering
\begin{tabular}{c|cc|ccc|ccc|ccc}
& \multicolumn{2}{c|}{Exp. C93} & \multicolumn{3}{c|}{Theory N05} & \multicolumn{3}{c|}{Num. K09} & \multicolumn{3}{c}{Present Num.}\\
 & $f_E$ $\left(\text{Hz}\right)$ & $\Delta_E$ $\left(\text{mHz}\right)$ & $f_T$ & $\Delta_T$ & $\Delta_E/\Delta_T$ & $f_N$ & $\Delta_N$ & $\Delta_E/\Delta_N$ & $f$ & $\Delta$ & $\Delta_E/\Delta$\\ \hline
 & 3.222 & 15$\pm$2 & 3.222 & 14.65 & 0.9767 & 3.228 & 16.27 & 1.0847 & 3.228 & 15.42 & 1.0267\\
\end{tabular}
\caption{Dimensional frequency and damping of the first non-axisymmetric mode $\left(1,1\right)$. Parameter setting: $R=0.05025\,\text{m}$, $h=0.13\,\text{m}$, $\rho=1000\,\text{kgm}^{-3}$, $\mu=0.00099\,\text{kgm}^{-1}\text{s}^{-1}$, $\gamma=0.0724\,\text{Nm}^{-1}$ and $\theta_s=62^{\circ}$, for which $Re=35\,628.103$ and $Bo=346.363$. Here $f=\left(\omega/2\pi\right)\sqrt{g/R}$ and $\Delta=\sigma\sqrt{g/R}$. The number of points in the radial and axial directions for the GLC grid used is this calculation is $N_r=N_z=40$, for which convergence is achieved.}
\label{tab:expC}
\end{table}

\subsubsection{Experiments by PD07 and theory by N05}\label{subsubsec:Sec4subsec4subsubsec3}

\begin{figure}
    \centering
     \includegraphics[width=1.\textwidth]{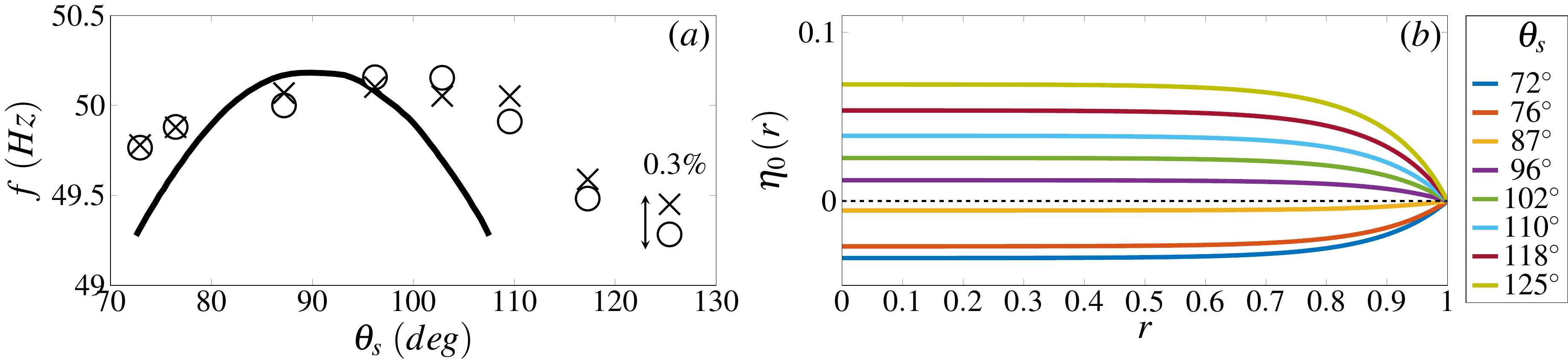}
    \caption{(a) Comparison of the experimentally measured natural frequency for mode $\left(0,10\right)$ (filled white circles, extracted from figure~5 of PD07) versus static contact angle with the inviscid estimation of N05 (black solid line) and our numerical results (black crosses). (b) Shape of the static meniscus, computed by solving Eq.~\eqref{eq:static_men}, corresponding to the contact angle values in (a). The black dashed line indicates the flat case with $\theta_s=90^{\circ}$. Parameter setting: pure water, clean surface, $h=0.045\,m$ and $R=0.025\,m$, for which $h/R=1.8$, $Bo=86.3$ and $Re=10\,855$. The number of points in the radial and axial directions for the GLC grid used is this calculation is $N_r=N_z=40$, for which convergence is achieved.}
    \label{fig:expPicard}
\end{figure}

PD07 presented a liquid surface biosensor for DNA detection based on resonant meniscus capillary waves. In their experimental setting the contact line is pinned at the brim, so that the static contact angle can be modified by controlling the bulk volume. As their setup was developed to make use exclusively of axisymmetric stationary meniscus waves, by exciting the container below the Faraday threshold they could measured the amplitude spectra for a series of effective contact angles in a frequency window centered around one particular natural frequency (that of mode $\left(m,n\right)=\left(0,10\right)$), enlightening two main phenomena attributable to contact angle effects, namely a decrease of the resonance frequency and a strong increase of the wave amplitude with the curvature of the meniscus, the latter being typical of a meniscus waves response. The experimental values were found to be in qualitative agreement with the inviscid prediction of N05, according to which the frequency has a maximum for $\theta_s=90^{\circ}$ (the maximum experimental frequency is found for $\theta_s\in\left[90,100\right]$). The frequency shift as a function of the static contact angle measured by PD07 is shown in figure.~\ref{fig:expPicard}(a) together with our numerical prediction for this specific case. Figure~\ref{fig:expPicard}(b) shows the shape of the static meniscus for the static contact angles (computed numerically by solving~\eqref{eq:static_men}) reported in figure~\ref{fig:expPicard}(a). Even in this case, our frequency prediction lies within 0.3\% the experimental values.

\subsubsection{Numerical study by K09}\label{subsubsec:Sec4subsec4subsubsec2}

\begin{figure}
    \centering
    \includegraphics[width=1.\textwidth]{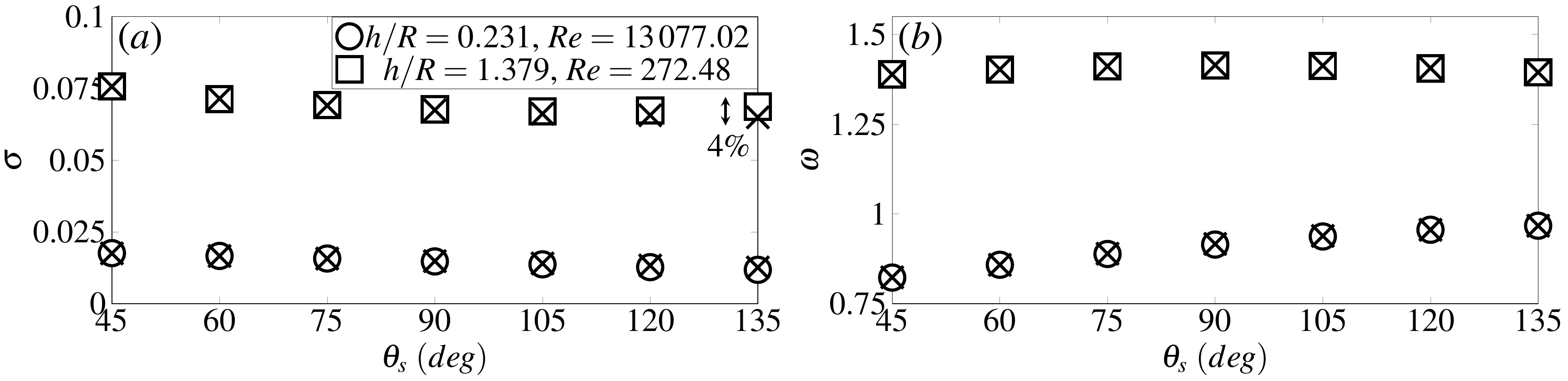}\\
    \bigskip
    \includegraphics[width=1.\textwidth]{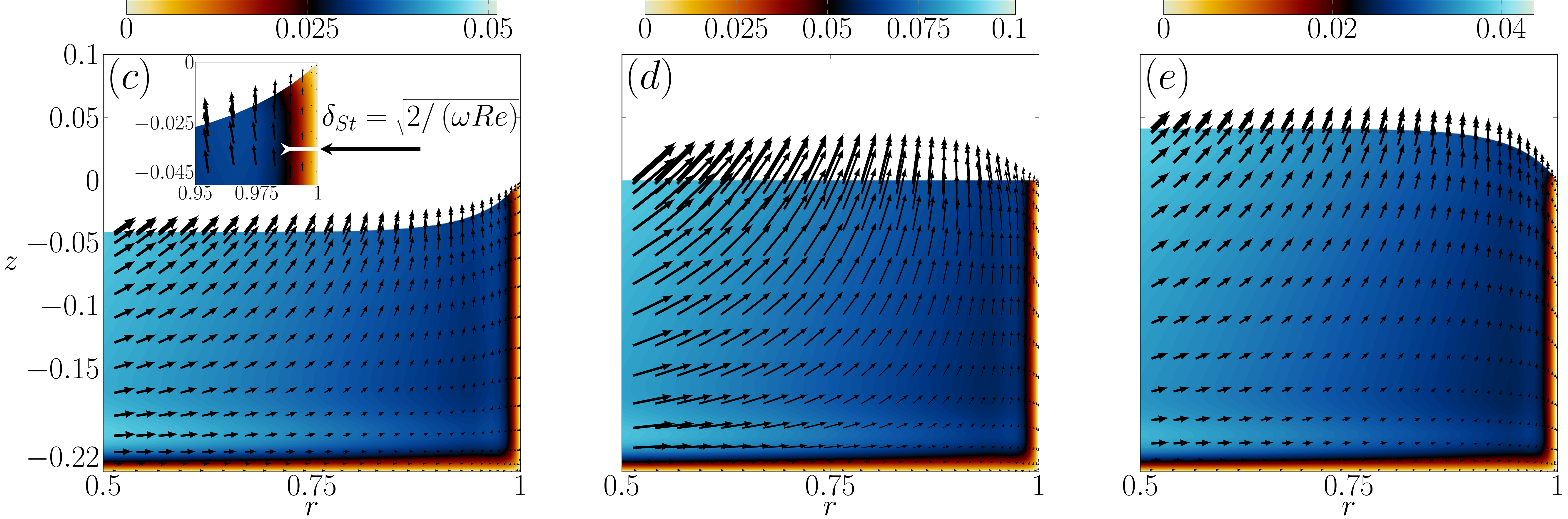}
    \caption{(a) Damping and (b) frequency of the first asymmetric mode $\left(1,1\right)$ as a function of the static contact angle. White filled squares and circles: numerical results of K09. Black crosses: present numerical results. The Bond number is fixed to $Bo=365$. The number of points in the radial and axial directions for the GLC grid used is this calculation is $N_r=N_z=40$, for which convergence is achieved. (c) Eigen-velocity field for $h/R=H=0.231$, $Re=13077.02$ and $\theta_s=45^{\circ}$ at $t=$ and $\phi=0$. Color plot: magnitude of the azimuthal eigen-velocity field, $|\hat{u}_{1\phi}|$. Arrows: imaginary parts of the in-plane eigen-velocity fields, Im$\left(\hat{u}_{1r}\right)$ and Im$\left(\hat{u}_{1z}\right)$. Only half domain is shown, $r\in\left[0.5,1\right]$. Inset: zoom on the meniscus region. (d)-(e) Same as (c) with $\theta_s=90^{\circ}$ and $\theta_s=135^{\circ}$, respectively. The eigenmodes are normalized with the phase and absolute value of the slope at the contact line, such that at large $Re$ the eigen-interface is predominantly real, whereas the eigen-wave velocity field is predominantly imaginary (the phase-shift between velocity and elevation is approximatively $\pi$/2).}
    \label{fig:numKidambi}
\end{figure}

As mentioned in the introduction, an important theoretico-numerical work accounting for contact angle effects on the damping and frequency of viscous capillary--gravity waves is that of K09. In figure~\ref{fig:numKidambi}(a) our predictions are compared with Kidambi's results for the first non-axisymmetric mode $\left(1,1\right)$ and for two different combination of nondimensional physical parameters. Our solution is found to be in good agreement with that of K09 for a wide range of static contact angle. In particular, the predicted frequencies are within 0.4\% with each other. Different peculiar behaviours are observed as the contact angle and the other physical parameters are varied. K09 found that at shallow depths the presence of a static meniscus leads to an increase of the natural frequency irrespective of the static contact angle, while at large depths the frequency shows a maximum in the neighborhood of $\theta_s=90^{\circ}$, in agreement with N05, with the experimental observations pointed out by PD07, and with the present study.

For completeness, the eigen-velocity field (solution of the eigenvalue problem~\eqref{eq:eig_eps1}) corresponding to the case of figure~\ref{fig:numKidambi}(a) and (b) with $h=0.231$ and $Re=13\,077.02$ and for three different static contact angles, is shown in figures~\ref{fig:numKidambi}(c), (d) and (e), from which the oscillating boundary layers at the solid bottom and lateral walls, whose thicknesses coincide precisely with the thickness of the Stokes boundary layer, $\delta_{St}=\sqrt{2/\left(\omega Re\right)}$ (as indicated in the inset of figure~\ref{fig:numKidambi}(c)), are clearly visible. 

\subsection{Comments}\label{subsec:Sec4subsec5}

Although the frequency predictions are in excellent agreement with experimental measurements (usually well within 1\%), we observe that the estimation of the damping coefficient is more sensitive to the various methods of calculation proposed in the literature. This is due to the fact that most of the existing theories are based on semi-analytical asymptotic expressions and boundary layer approximations with a leading order solution formulated in the inviscid framework (HM94,M98,MH98,N02,N05), as originally introduced by \cite{Benjamin79} and \cite{graham1983new}. However, despite the sources of dissipation may be several and hard to accurately quantify, especially with asymptotic approaches, the pinned contact line problem allows one to drastically reduce uncertainties related to contact line dynamics, thus leading in general to better agreements with experiments. Little uncertainties can still be present in experiments, where free surface contamination is not fully controlled.\\
\indent A wide majority of studies, both experimental and numerical (or semi-analytic), have been focused on the classic case of a flat static free surface, with the exceptions of those by N05 and K09. Particularly K09, in the spirit of N02, projected the governing equations onto an appropriate basis and formulated a nonlinear eigenvalue problem (solved numerically with an iterative method) for the damping and frequency of viscous capillary--gravity waves with fixed contact line, which formally includes both static meniscus effects and viscous dissipation.\\
\indent We underly that, unlike the previous analyses by K09, in the present work, through a fully numerical discretization technique, the problem of viscous capillary--gravity waves with pinned contact line is formulated as a classic generalized linear eigenvalue problem, which can be solved numerically with standard techniques. The damping coefficients and natural frequencies have been shown in this section to be in fairly good agreement with previous experiments (for both $\theta_s=90^{\circ}$ and $\theta_s\ne90^{\circ}$). Moreover, the spatial structures of the perturbation wave fields are found as the eigenvalues and eigenmodes, respectively, of the 2-dimensional (in $r$-$z$) linear problem~\eqref{eq:eig_eps1}. In other words, all the information associated with the $\left(m,n\right)$ wave is contained in the complex eigenfunction, $\hat{\mathbf{q}}_1$ and $\hat{\eta}_1$, which satisfies the prescribed boundary conditions by construction, and in its corresponding complex eigenvalue, $\lambda=-\sigma+\text{i}\omega$. Furthermore, the numerical method used in this work allows one to directly solve capillary--gravity waves in perfect and nearly--brimful condition accounting for contact angle effects and viscous dissipation without any simplification or assumption, i.e. the numerical solution at convergence is supposed to be accurate.\\


\section{Weakly--nonlinear model for sub--harmonic Faraday thresholds with contact angle effects}\label{sec:Sec5}

In this section, the numerical tools presented and validated in \S\ref{sec:Sec4} are employed to formalize a weakly nonlinear model accounting for contact angle effects, i.e. static meniscus and harmonic meniscus capillary waves, on the sub--harmonic Faraday instability with pinned contact line.

\subsection{Presentation}\label{subsec:Sec5subsec1}

Here, the full system~\eqref{eq:GovEqNS}-\eqref{eq:PinnBC} is solved through a weakly nonlinear (WNL) analysis based on the multiple scale method that is valid in the regime of small perturbations of the static configuration and small external control parameters, namely the driving forcing amplitude and detuning from the parametric resonance. Let us thus introduce the following asymptotic expansion for the flow quantities,

\begin{equation}
\label{eq:WNLexp_q}
\mathbf{q}=\left\{\mathbf{u},p\right\}^T=\mathbf{q}_0+\epsilon\mathbf{q}_1+\epsilon^2\mathbf{q}_2+\epsilon^3\mathbf{q}_3+\text{O}\left(\epsilon^4\right),
\end{equation}
\begin{equation}
\label{eq:WNLexp_eta}
\eta=\eta_0+\epsilon\eta_1+\epsilon^2\eta_2+\epsilon^3\eta_3+\text{O}\left(\epsilon^4\right).
\end{equation}

\noindent In the spirit of the multiple scale technique, we introduce the slow time scale $T=\epsilon^2 t$, with $t$ being the fast time scale at which the free surface oscillates. Since we focus on sub--harmonic resonances, the system is expected to respond with a frequency equal to half the driving frequency, therefore we assume the external forcing angular frequency to be $\Omega_d=2\omega+\Lambda$, where $\omega$ is the natural frequency associated with the generic $\left(m,n\right)$ capillary--gravity wave considered and $\Lambda$ is the detuning parameter. As, by construction, the WNL analysis is valid close to the instability threshold only, we assume a departure from criticality to be of order $\epsilon^2$. In terms of control parameters, this assumption translates in the following scalings for the external forcing amplitude, $F_d/g$, and detuning $\Lambda$, 
\begin{equation}
\label{eq:WNLexp_forc_amp_freq}
F_d/g=F=\epsilon^2\hat{F},\ \ \ \ \ \Lambda=\epsilon^2\hat{\Lambda}.
\end{equation}
\noindent It should be noted that the presence of viscosity leads to a damped $\epsilon$--order solution $\mathbf{q}_1$ (as discussed in \S\ref{sec:Sec4}), whereas standard multiple scale methods apply to marginally stable systems \citep{Nayfeh2008}. Nevertheless, as the Reynolds is typically high enough, the damping coefficient results in a slow damping process over fast wave oscillations (see \S\ref{sec:Sec4}). In such a regime, a multiple scale analysis can still be applied by postulating that the damping coefficient of the $\left(m,n\right)$ wave is of order $\epsilon^2$, i.e. $\sigma=\epsilon^2\hat{\sigma}$, therefore the $\left(m,n\right)$ eigenvalue reads $\lambda=-\epsilon^2\hat{\sigma}+\text{i}\omega$. A simple way to account for this second order departure from neutrality consists in replacing the leading order operator $\mathcal{A}_m=\mathcal{A}_m\left(Re\right)$ defined in~\eqref{eq:eig_eps1}, for which $\hat{\mathbf{q}}_1$ is not neutral, but rather stable, by the \textit{shifted} operator \citep{Meliga2009}, $\tilde{\mathcal{A}}_m=\mathcal{A}_m+\epsilon^2\mathcal{S}_m$, where $\mathcal{S}_m$ is the shift operator defined as $\mathcal{S}_m\hat{\mathbf{q}}_1=-\hat{\sigma}\hat{\mathbf{q}}_1$. The shifted operator $\tilde{\mathcal{A}}_m$ is characterized by the same spectra of $\mathcal{A}_m$, excepted that the $\left(m,n\right)$ eigenmode $\hat{\mathbf{q}}_1$ associated with $\hat{\sigma}$ is now marginally stable, and hence the WNL formalism can be applied. For a thorough discussion about the formalism of the shift operator see \cite{Meliga2009,Meliga2012}. Although a different approach to account for a damped first order solution was followed by \cite{Viola2018b}, leading to a different (but equivalent) asymptotic expansion, we use in this paper the shift operator approach.\\
\indent Finally, substituting the asymptotic expansions and scalings above in the governing equations~\eqref{eq:GovEqNS}-\eqref{eq:PinnBC} with their boundary conditions, a series of problems at the different orders in $\epsilon$ are obtained.\\
\indent As anticipated in \S\ref{sec:Sec3}, when contact angle effects are included in the analysis, i.e. the initial static interface is not flat, the third order asymptotic expansion of the full viscous hydrodynamic system introduced in \S\ref{sec:Sec2} turns out to be very complex to be derived analytically. Particularly tedious is the dynamic boundary condition, as it involves free surface boundary terms, which, within the linearization process, must be flattened at the static interface, $\eta_0$, as well as the full nonlinear curvature. In order to overcome these practical difficulties, the linearization and expansion procedures have been fully automated using symbolic calculus within the software Wolfram Mathematica, which has been then integrated within the main code implemented in Matlab. The corresponding Mathematica codes are provided as supplementary material.

\subsection{Order $\epsilon^0$: static meniscus}\label{subsec:Sec5subsec2}
At order $\epsilon^0$ the system reduces to the nonlinear equation associated with the shape of the axisymmetric static meniscus. The velocity field is null, $\mathbf{u}_0=\mathbf{0}$ and the pressure is hydrostatic, $p_0=-z$. As described in \S\ref{subsec:Sec4subsec1}, the static interface, $\eta_0\left(r\right)$, is obtained by prescribing a static contact angle, $\theta_s$, which enters through the geometrical relation~\eqref{eq:static_men_geom_rel} imposed at the contact line.

\subsection{Order $\epsilon$: capillary--gravity waves}\label{subsec:Sec5subsec3}
At leading order in $\epsilon$ the system is represented by the unsteady Stokes equations~\eqref{eq:GovEqNS_eps1}, together with the kinematic and dynamic boundary conditions~\eqref{eq:bcKin_eps1}-\eqref{eq:bcDyn_eps1}, linearized around the static base flow $\mathbf{q}_0=\left\{\mathbf{u}_0,p_0\right\}^T=\left\{\mathbf{0},-z\right\}^T$ and $\eta_0$, and subjected to the no-slip b.c. at the solid walls, regularity conditions at the axis~\eqref{eq:regularity_cond_m0}-\eqref{eq:regularity_cond_m234}, and to the pinned contact line condition~\eqref{eq:BcCLpinn}:
\begin{equation}
\label{eq:WNLeps1_sys}
\left(\mathcal{B}\partial_t-\tilde{\mathcal{A}}_m\right)\mathbf{q}_1=\mathbf{0}.
\end{equation}

\noindent Within the framework of the Faraday instability, we are interested in a standing wave form of the solution, which can be seen as a results of the balance of two counter rotating waves. Hence, we seek for a first order solution of the form
\begin{eqnarray}
\label{eq:q1eps}
\boldsymbol{q}_1=A_1^+\left(T\right)\hat{\boldsymbol{q}}_1^{A^+}e^{\text{i}\left(\omega t+m\phi\right)}+A_1^-\left(T\right)\hat{\boldsymbol{q}}_1^{A^-}e^{\text{i}\left(\omega t-m\phi\right)}+c.c.,
\end{eqnarray}

\noindent that destabilizes the static configuration. A single azimuthal wavenumber $m$ is considered at a time. In~\eqref{eq:q1eps} $A^+_1$ and $A^-_1$, unknown at this stage of the expansion, are the complex amplitudes of the oscillating mode $\hat{\mathbf{q}}_1^{A^+}$ and $\hat{\mathbf{q}}_1^{A^-}$ respectively and they are function of the slow time scale $T$. The eigensolution of~\eqref{eq:WNLeps1_sys} has been widely discussed in \S\ref{sec:Sec4} for $m\geqslant0$. We note in addition that the eigenmode for the $-m$ perturbation is similar to that of the $+m$ perturbation, more precisely, it oscillates with the same frequency $\omega$, but it has the opposite pitch and during time it rotates in the opposite direction.

\subsection{Order $\epsilon^2$: meniscus waves, second-harmonics and mean-flow corrections}\label{subsec:Sec5subsec4}

At order $\epsilon^2$ we obtain the linearized Stokes equations and boundary conditions applied to $\mathbf{q}_2=\left\{\mathbf{u}_2,p_2\right\}^T$ and $\eta_2$,
\begin{equation}
\label{eq:eps2_compact}
\left(\mathcal{B}\partial_t-\tilde{\mathcal{A}}_m\right)\mathbf{q}_2=\boldsymbol{\mathcal{F}}_2,
\end{equation}

\noindent and forced by a term $\boldsymbol{\mathcal{F}}_2$ depending only on zero-, first-order solutions and on the external forcing

\begin{eqnarray}
\label{eq:q2epsF}
\boldsymbol{\mathcal{F}}_2=|A^+_1|^2\hat{\boldsymbol{\mathcal{F}}}_2^{A^+A^{+^*}}+|A^-_1|^2\hat{\boldsymbol{\mathcal{F}}}_2^{A^-A^{-^*}}+\left(\hat{F}\hat{\boldsymbol{\mathcal{F}}}_2^{\hat{F}}e^{\text{i}\left(2\omega t+\hat{\Lambda}T\right)}+c.c.\right)+\\
+\left(A^{+^2}_1\hat{\boldsymbol{\mathcal{F}}}_2^{A^+A^+}e^{\text{i}\left(2\omega t+2m\phi\right)}+A^{-^2}_1\hat{\boldsymbol{\mathcal{F}}}_2^{A^-A^-}e^{\text{i}\left(2\omega t-2m\phi\right)}+c.c.\right)+\notag\\
+\left(A^+_1A^-_1\hat{\boldsymbol{\mathcal{F}}}_2^{A^+A^-}e^{\text{i}2\omega t}+A^+_1A^{-^*}_1\hat{\boldsymbol{\mathcal{F}}}_2^{A^+A^{-^*}}e^{\text{i}2m\phi}+c.c.\right)\notag.
\end{eqnarray}

\noindent All terms contributing to the forcing vector $\boldsymbol{\mathcal{F}}_2$ were extracted using symbolic calculus in Wolfram Mathematica (see supplementary material). The first order solution is made of four different contributions of amplitude $A^+_1$, $A^{+^*}_1$, $A^-_1$ and $A^{-^*}_1$, therefore it generates 10 different second order forcing terms, $\hat{\boldsymbol{\mathcal{F}}}_2^{ij}e^{\text{i}\left(\omega^{ij} t+m^{ij}\phi\right)}$, which exhibits a certain frequency and spatial periodicity, gathered in table~\ref{tab:conv_normal_form}. The two additional terms, $\hat{\boldsymbol{\mathcal{F}}}_2^{\hat{F}}$, appearing in the forcing expression~\eqref{eq:q2epsF}, comes from the spatially uniform axisymmetric external forcing typical of Faraday waves, whose amplitude was assumed to be of order $\epsilon^2$.

\begin{table}
\centering
\begin{tabular}{c|ccccccc}
$\epsilon^2$ & $A^+_1A^{+*}_1$ & $A^-_1A^{-*}_1$ & $\hat{F}$ & $A^+_1A^+_1$ & $A^-_1A^-_1$ & $A^+_1A^-_1$ & $A^+_1A^{-*}_1$\\ \hline
$m^{ij}$ & 0 & 0 & 0 & 2$m$ & -2$m$ & 0 & 2$m$\\
$\omega^{ij}$ & 0 & 0 & 2$\omega$ & 2$\omega$ & 2$\omega$  & 2$\omega$ & 0\\
\end{tabular}
\caption{Second order nonlinear forcing terms gathered by their amplitude dependency, and corresponding azimuthal and temporal periodicity $\left(m^{ij},\omega^{ij}\right)$. Seven terms have been omitted as they are the complex conjugates.}
\label{tab:conv_normal_form}
\end{table}

\noindent All these forcing terms are non-resonant, as their oscillation frequencies and their spatial symmetries, through the azimuthal wavenumber, differ from those of the leading order solution (see table~\ref{tab:conv_normal_form}). Hence no solvability conditions are required at the present order \citep{Meliga2009}. We can thus seek for a second order solution as the superimposition of the second order response to the external forcing, $\hat{\mathbf{q}}_2^{\hat{F}}$, and 10 responses $\hat{\mathbf{q}}_2^{ij}$ to each single forcing terms, 

\begin{eqnarray}
\label{eq:q2eps}
\boldsymbol{q}_2=|A^+_1|^2\hat{\mathbf{q}}_2^{A^+A^{+^*}}+|A^-_1|^2\hat{\mathbf{q}}_2^{A^-A^{-^*}}+\left(\hat{F}\hat{\mathbf{q}}_2^{\hat{F}}e^{\text{i}\left(2\omega t+\hat{\Lambda}T\right)}+c.c.\right)+\\
+\left(A^{+^2}_1\hat{\mathbf{q}}_2^{A^{+^2}}e^{\text{i}\left(2\omega t+2m\phi\right)}+A^{-^2}_1\hat{\mathbf{q}}_2^{A^{-^2}}e^{\text{i}\left(2\omega t-2m\phi\right)}+c.c.\right)+\notag\\
+\left(A^+_1A^-_1\hat{\mathbf{q}}_2^{A^+A^-}e^{\text{i}2\omega t}+A^+_1A^{-^*}_1\hat{\mathbf{q}}_2^{A^+A^{-^*}}e^{\text{i}2m\phi}+c.c.\right)\notag,
\end{eqnarray}

\begin{figure}
    \centering
    \includegraphics[width=\textwidth]{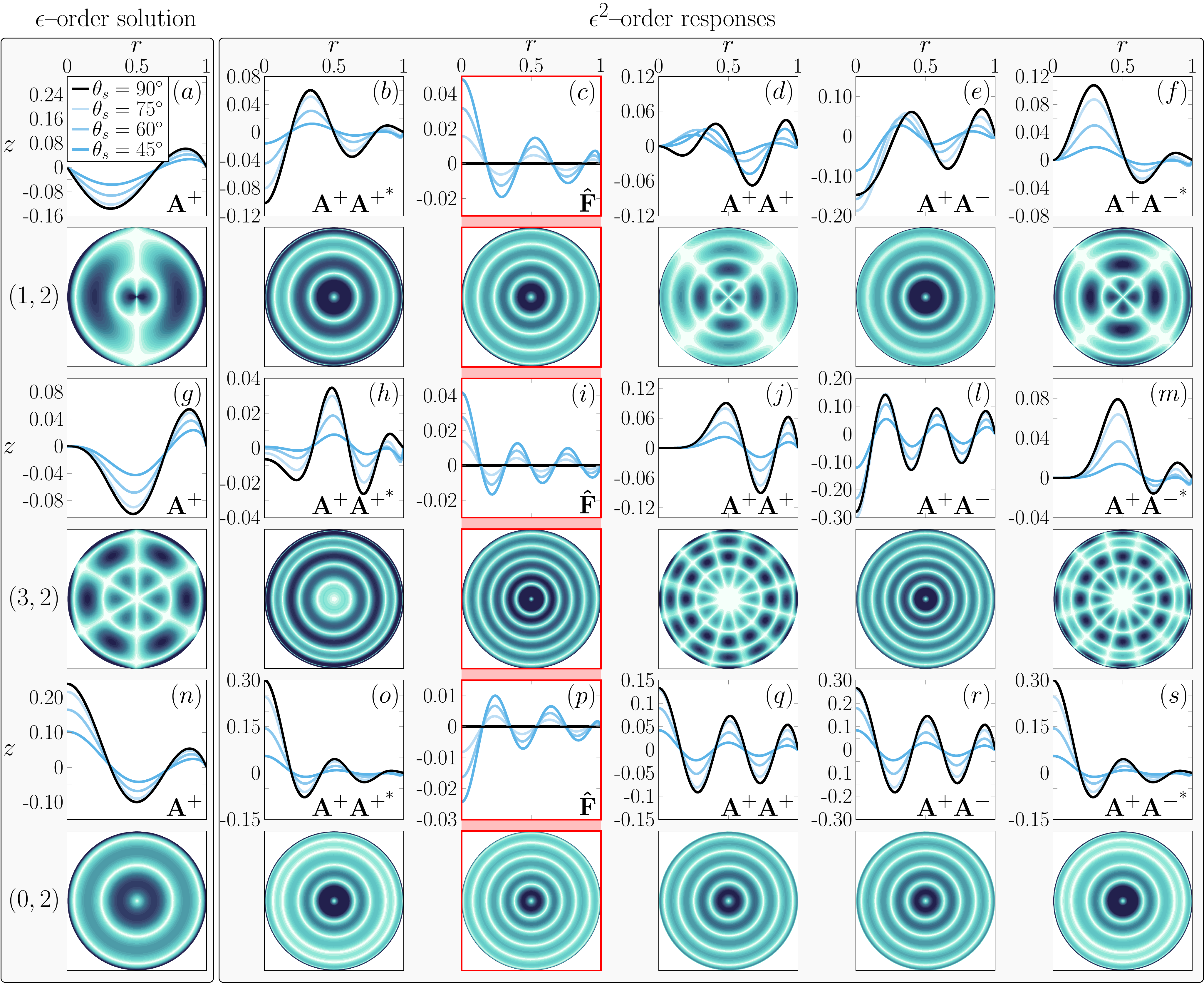}
    \caption{(a)-(f) Top: real part of the free surface elevation, Re$\left(\hat{\eta}\right)$ associated with (a) mode $\left(1,2\right)$ and with (b)-(f) some of the corresponding second order responses for different values of the static contact angle, $\theta_s$. The $\epsilon$--order solution is normalized with the absolute value and phase of the contact line slope. Bottom: free surface visualization in terms of absolute value of the real part of the interface slope at $\theta_s=45^{\circ}$. The colormaps were individually saturated for visualization purposes only. (g)-(m) Same as (a)-(f), but for mode $\left(3,2\right)$. (n)-(s) Same as (a)-(f), but for the axisymmetric mode $\left(0,2\right)$. Parameter setting: $R=0.035\,\text{m}$, $h=0.022\,\text{m}$, $\rho=997\,\text{kgm$^{-3}$}$, $\mu=0.001\,\text{kgm$^{-1}$s$^{-1}$}$, $\gamma=0.072\,\text{Nm$^{-1}$}$, for which $Bo=166.2$ and $Re=20\,437$, and a static contact angle $\theta_s=45^{\circ}$. The light red boxes highlights the second order response to the external forcing, i.e. second order harmonic meniscus waves.}
    \label{fig:eps1_eps2_sol_surf_mn32_mn02}
\end{figure}

\noindent each of which is computed as a solution of a linear forced problem

\begin{equation}
\label{eq:WNLeps2_lin_forc_sys}
\left(\text{i}\omega^{ij}\mathcal{B}-\tilde{\mathcal{A}}_{m^{ij}}\right)\hat{\mathbf{q}}_2^{ij}=\hat{\boldsymbol{\mathcal{F}}}_2^{ij},
\end{equation}

\noindent with $m^{ij}$ and $\omega^{ij}$ for $\left(i,j\right)$ from table~\ref{tab:conv_normal_form} and which can be inverted (non-singular operator) as long as any of the combinations $\left(m^{ij},\omega^{ij}\right)$ is not an eigenvalue (none of them has $m^{ij}=\pm m$). As an example, the $\epsilon$--order eigensurface and some of the various second order surfaces are shown in figure~\ref{fig:eps1_eps2_sol_surf_mn32_mn02} for three different waves, i.e. $\left(m,n\right)=\left(1,2\right)$, $\left(3,2\right)$ and $\left(0,2\right)$. Owing to the symmetries of the system (given in equation~\eqref{eq:m_transf}), some of the second order responses corresponding to the generic $\left(m,n\right)$ wave have the same solution with opposite azimuthal velocity, therefore in figure~\ref{fig:eps1_eps2_sol_surf_mn32_mn02} we show only the solutions with different surface shapes. Furthermore, as can be deduced from figure~\ref{fig:eps1_eps2_sol_surf_mn32_mn02}(n)-(s), in the axisymmetric case $\left(0,n\right)$ all the responses are axisymmetric with zero azimuthal velocity, thus some of the second order responses share exactly the same solution. In this case, indeed, the second order solution could be formulated \textit{a priori} as the sum of three terms only, whose amplitudes are proportional to $\hat{F}$, $A^2_1$ (second harmonic) and $|A_1|^2$ (mean flow correction), respectively.\\
\indent Of particular interest is the second order response to the external forcing, whose interface shape is highlighted by the red boxes in figure~\ref{fig:eps1_eps2_sol_surf_mn32_mn02}. With the present scaling, the forcing enters at second order in the $z$--component of the momentum equation (see equation~\eqref{eq:GovEqNS}). If the initial static interface is assumed to be flat ($\theta_s=90^{\circ}$), then the response $\left(\hat{\mathbf{q}}_2^{\hat{F}},\hat{\eta}_2^{\hat{F}}\right)$, translates into a harmonic hydrostatic pressure modulation only, with a free surface remaining flat, i.e. $\hat{\mathbf{u}}_2^{\hat{F}}=\mathbf{0}$ and $\hat{\eta}_2^{\hat{F}}=0$, a case classically analyzed in literature. On the other hand, as shown in figure~\ref{fig:eps1_eps2_sol_surf_mn32_mn02}(c), (i) and (p), if a static contact angle $\theta_s\ne90^{\circ}$ is considered, then the $\epsilon^0$--order static meniscus induces at order $\epsilon^2$ axisymmetric meniscus capillary waves traveling from the sidewall to the interior and reflected back, which oscillates harmonically with the external forcing and with an amplitude proportional to the external forcing amplitude. In the present WNL analysis, these meniscus waves, which appears as concentric ripples (see figure~\ref{fig:eps1_eps2_sol_surf_mn32_mn02}(c), (i) and (p)), as typically observed in experiments \citep{batson2013faraday,shao2021role,shao2021surface}, will couple at third order with the first order solution and will contribute to modify both the linear stability boundaries associated with the sub--harmonic Faraday tongues as well as the bifurcation diagram, i.e. wave amplitude saturation to finite amplitude. Furthermore, figure~\ref{fig:eps1_eps2_sol_surf_mn32_mn02} clearly shows that a static contact angle $\theta_s\ne90$, depending on its value (here only values of $\theta_s<90^{\circ}$ have been considered), modifies not only the damping coefficients and frequencies of the leading order wave (see also figure~\ref{fig:expPicard} and~\ref{fig:numKidambi}), but also its spatial shape and, as consequence, all the associated second order responses, whose modifications may have a significant influence on the corresponding saturation to a finite amplitude. 

\subsection{Order $\epsilon^3$: amplitude equation for standing waves}\label{subsec:Sec5subsec5}

Lastly, at the $\epsilon^3$--order we derive an amplitude equation for standing waves with a pinned contact line accounting for weakly nonlinear modifications of the sub--harmonic Faraday threshold owing to contact angle effects. The problem at order $\epsilon^3$ is similar to the one obtained at order $\epsilon^2$, as it appears as a linear system,

\begin{equation}
\label{eq:eps3_compact}
\left(\mathcal{B}\partial_t-\tilde{\mathcal{A}}_m\right)\mathbf{q}_3=\boldsymbol{\mathcal{F}}_3,
\end{equation}

\noindent forced by a combinations of the previous order solutions englobed in $\boldsymbol{\mathcal{F}}_3$, that contains several nonlinear terms of various space and time periodicities and which we denote as $\hat{\boldsymbol{\mathcal{F}}}_3^{ij}e^{\text{i}\left(\omega t + m\phi \right)}$. Since many of these terms are resonant, as standard in multiple scale analysis, in order to avoid secular terms and solve the expansion procedure at the third order, a compatibility condition must be enforced through the Fredholm alternative \citep{friedrichs2012spectral}. Such a compatibility condition imposes the amplitudes $A^+_1$ and $A^-_1$ to obey the following relation
\begin{equation}
\label{eq:eps3AmpEq_pm}
\frac{dA^{\pm}}{dt}=-\sigma A^{\pm} + \zeta FA^{{\mp}^*}e^{\text{i}\Lambda t / 2} + \chi_1 |A^{\pm}|^2A^{\pm} + \chi_2 |A^{\mp}|^2A^{\pm},
\end{equation}

\noindent where the physical time $t=T/\epsilon^2$ has been reintroduced and where $\sigma=\epsilon^2\hat{\sigma}$, $F=F_d/g=\epsilon^2\hat{F}$ and $\Lambda=\epsilon^2\hat{\Lambda}$. By considering the expansion $\mathbf{q}=\mathbf{q}_0+\epsilon A_1\hat{\mathbf{q}}_1\hdots$, the small parameter $\epsilon$ is eliminated by defining the amplitude $A=\epsilon A_1$, so that everything is recast in terms of actual physical quantities. The various coefficients are computed as scalar product between the adjoint global modes and the resonant forcing terms $\hat{\boldsymbol{\mathcal{F}}}_3^{ij}$, whose analytically complex expressions have been extracted from the third order forcing using the symbolic calculus tools of Wolfram Mathematica. For instance, the complex coefficient $\zeta$ is evaluated as
\begin{equation}
\label{eq:WNLeps3_ex_coeff}
\small \zeta=\frac{\int_{\text{V}}\hat{\mathbf{u}}_1^{\dagger^* A^{+}}\cdot\hat{\boldsymbol{\mathcal{F}}}_{3,Mom}^{\hat{F}A^{-^*}}\,r\text{d}r\text{d}z+\int_{\eta_0}\hat{\mathbf{u}}_1^{\dagger^* A^{+}}\cdot\hat{\boldsymbol{\mathcal{F}}}_{3,Dyn}^{\hat{F}A^{-^*}}\,r\text{d}r+\int_{\eta_0}\xi^{\dagger^* A^+}\hat{\mathcal{F}}_{3,Kin}^{\hat{F}A^{-^*}}\,r\text{d}r}{\int_{\text{V}}\hat{\mathbf{u}}_1^{\dagger^* A^{+}}\cdot\hat{\mathbf{u}}_1^{A^{+}}\,r\text{d}r\text{d}z+\int_{\eta_0}\xi^{\dagger^* A^+}\hat{\eta}_1^{A^+}\,r\text{d}r}
\end{equation}

\noindent where $\text{V}$ denotes the fluid bulk domain, the dagger symbol refers to the adjoint eigenmode, $\xi = \left[-\frac{\eta_{0,r}}{Re}\left(\frac{\partial \hat{u}_{1z}}{\partial r}+\frac{\partial \hat{u}_{1r}}{\partial z}\right)+\left(-\hat{p}_1+\frac{2}{Re}\frac{\partial \hat{u}_{1z}}{\partial z}\right)\right]$ (see also \cite{Viola2018b}) and the subscripts $_{NS}$, $_{Dyn}$ and $_{Kin}$ designate the forcing components of $\hat{\boldsymbol{\mathcal{F}}}_3^{\hat{F}A^{-^*}}$ appearing in the $\epsilon^3$--order momentum equations, dynamic boundary condition and kinematic boundary condition, respectively. Analogous expressions hold for $\chi_1$ and $\chi_2$ by replacing $\hat{\boldsymbol{\mathcal{F}}}_3^{\hat{F}A^{-^*}}$ with $\hat{\boldsymbol{\mathcal{F}}}_3^{A^+A^{+^*}A^+}$ and $\hat{\boldsymbol{\mathcal{F}}}_3^{A^-A^{-^*}A^+}$, respectively. We notice that the adjoint eigenvector appearing in~\eqref{eq:WNLeps3_ex_coeff} does not need to be independently calculated. Indeed, \cite{Viola2018b} demonstrated that the linear operator $\mathcal{B}$ and $\mathcal{A}_m$ (the same applies to the shifted operator $\tilde{\mathcal{A}}_m$) are self-adjoint, i.e. $\mathcal{B}^{\dagger}=\mathcal{B}$ and $\mathcal{A}_m^{\dagger}=\mathcal{A}_m$, with the adjoint eigenvalue being the complex conjugate of the direct one, $\lambda^{\dagger}=\lambda^*$. Then, from~\eqref{eq:m_transf}, \eqref{eq:omega_transf1} and \eqref{eq:omega_transf2}, it follows that for the couple $\left(m,-\sigma+\text{i}\omega\right)$ associated with a direct mode, we have the relation
\begin{equation}
\label{eq:WNLeps3_adj}
\left(\hat{u}_{1r}^*,-\hat{u}_{1\phi}^*,\hat{u}_{1z}^*,\hat{p}_1^*,\hat{\eta}_1^*\right)\rightarrow\left(\hat{u}_{1r}^{\dagger},\hat{u}_{1\phi}^{\dagger},\hat{u}_{1z}^{\dagger},\hat{p}_1^{\dagger},\hat{\eta}_1^{\dagger}\right),
\end{equation}

\noindent which directly provide the desired adjoint mode without any further calculation. We also underly that due to the symmetry of the solution, the same value of $\zeta$ is obtained if one makes use of the scalar product between the adjoint mode for $A^-_1$ and the forcing term $\hat{\boldsymbol{\mathcal{F}}}_3^{\hat{F}A^{+^*}}$ (same for $\chi_1$ and $\chi_2$).\\
\indent As anticipated before, the standing wave solution corresponds to the superimposition of two balanced counter-rotating waves of same amplitude $A^+=A^-=A$. It follows that system~\eqref{eq:eps3AmpEq_pm} reduces to the single amplitude equation

\begin{equation}
\label{eq:eps3AmpEqFin}
\frac{dB}{dt}=-\left(\sigma+\text{i}\Lambda/2\right) B + \zeta FB^* + \chi |B|^2B ,
\end{equation}

\noindent where the change of variable $A=B^{\text{i}\Lambda/2}$ has been introduced and where the complex coefficient $\chi$ is taken as the sum of $\chi_1$ and $\chi_2$. The form of~\eqref{eq:eps3AmpEqFin} is totally equivalent to the normal form~\eqref{eq:amp_eq_intro} postulated by \cite{Douady90} using symmetry arguments only. Its structure indeed does not depend on the boundary conditions and on the mode shape, nevertheless its coefficients do. In the present work these complex coefficients, $\zeta$ and $\chi$, as well as the frequency and damping of the wave, $\omega$ and $\sigma$, are formally computed by taking into account the full hydrodynamic system, whose solution is exact at numerical convergence. The damping coefficient must be small enough, but its value is numerically computed, rather than estimated heuristically. Most importantly, $\zeta$ and $\chi$, through the WNL formulation presented above, englobe in a formal manner, although within the assumptions of validity of a single-mode WNL theory, the effect of the static contact angle and of the coupling with harmonic meniscus waves (MW) on the sub--harmonic Faraday threshold of standing viscous capillary--gravity waves with pinned contact line.

\subsection{Linear stability of the amplitude equation: sub-harmonic Faraday tongues}\label{subsec:Sec5subsec6}

Here we perform the stability analysis of the amplitude equation~\eqref{eq:eps3AmpEqFin}, which prescribes the marginal stability boundaries, typically known as Faraday tongues. By turning to polar coordinates 

\begin{equation}
\label{eq:WNLeps3_amp_phase_coeff}
B=|B|e^{\text{i}\Phi},\ \ \ \ \ -\left(\sigma+\text{i}\Lambda/2\right)=c_1e^{\text{i}\varphi_1},\ \ \ \ \ \zeta=c_2e^{\text{i}\varphi_2},\ \ \ \ \ \chi=c_3e^{\text{i}\varphi_3},
\end{equation}

\noindent splitting the modulus and phase parts of~\eqref{eq:eps3AmpEqFin} and introducing the change of variable $\Theta=\Phi-\varphi_2/2$, we obtain the following system

\begin{equation}
\label{eq:WNLeps3_ampB}
\frac{d|B|}{dt}=c_1\cos{\left(\varphi_1\right)}|B|+c_2\cos{\left(2\Theta\right)}F|B|+c_3\cos{\left(\varphi_3\right)}|B|^3,
\end{equation}
\begin{equation}
\label{eq:WNLeps3_phaseB}
\frac{d\Theta}{dt}=c_1\sin{\left(\varphi_1\right)}-c_2\sin{\left(2\Theta\right)}F+c_3\sin{\left(\varphi_3\right)}|B|^2.
\end{equation}

\noindent Equation~\eqref{eq:WNLeps3_ampB} admits two possible equilibria $\left(d/dt=0\right)$, having $|B|=0$ and $|B|\ne0$, respectively. We first focus on the stability of the trivial stationary solution, $|B|=0$. By eliminating $\Theta$ from~\eqref{eq:WNLeps3_ampB}-\eqref{eq:WNLeps3_phaseB}, the linear threshold or marginal stability boundaries (sub--harmonic Faraday tongues) are readily obtained \citep{Douady90,Rajchenbach2015},
\begin{equation}
\label{eq:WNLeps3_stab_bound}
F_{th}^L=\left(F_d/g\right)_{th}^L=c_1/c_2\ \longrightarrow\ F_{th}^L=\pm|\zeta|^{-1}\sqrt{\sigma^2+\left(\Omega_d/2-\omega\right)^2},
\end{equation}

\noindent where the relation $\Lambda=\Omega_d-2\omega$ has been reintroduced and which predicts the lowest threshold, $F_{th,min}^L=\sigma/|\zeta|$, at $\Omega_d=2\omega$. The forcing amplitude at which the mode appears is therefore proportional to its dissipation, $\sim\sigma$ (note that this is true only for sub--harmonic resonances, e.g. the threshold for harmonic tongues is expected to scale as $\sim\sigma^{1/2}$, see \cite{Rajchenbach2015}). Moreover, $F_{th}^L$ depends on the coefficient $\zeta$, which is produced by the interaction of the first order response, proportional to the amplitude $A^{\pm}$, with the second order response to the external forcing, proportional to $\hat{F}$. Therefore, contact angle modifications of the leading order solution and harmonic meniscus waves (see figure~\ref{fig:eps1_eps2_sol_surf_mn32_mn02}) enter directly in the calculation of $\zeta$, whose value contributes to the definition of the marginal stability boundaries. Presence of a static meniscus, as widely discussed in \S\ref{sec:Sec4}, also modify the natural frequency $\omega$ and the damping $\sigma$.

\subsubsection{Brimful condition: validation with the inviscid analysis by K13 for $\theta_s=90^{\circ}$}\label{subsubsec:Sec5subsec6subsubsec0}

\begin{figure}
    \centering
    \includegraphics[width=1.\textwidth]{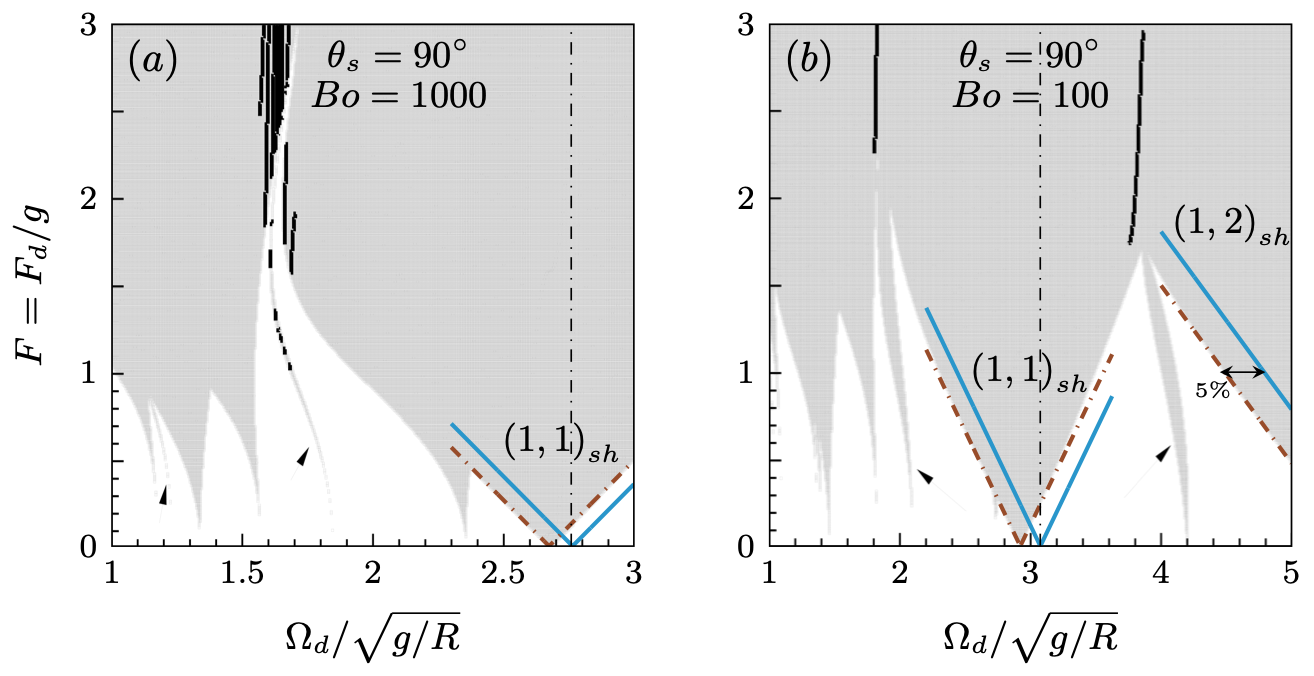}
    \caption{Inviscid stability plots associated with modes $\left(1,1\right)$ and $\left(1,2\right)$ for two different Bond numbers, i.e. $Bo=1000$ and $100$, and for a depth $h/R=H=1$. The gray shaded regions have not been reproduced in this work, but rather they have been simply taken from figure~10 of K13. Sub--harmonic tongues are denoted by the subscript $_{sh}$. For computational reasons, the instability  regions (gray shaded) were obtained in K13 by truncating the number of basis function $N_{K13}$ to 2, although convergence of the natural frequencies was achieved by taking $N_{K13}=30$, as stated by K13 in his table~1 (with a systematic underestimation of approximately 5\%). The vertical black dash-dot lines corresponds to the converged results reported in table~1 of K13. The blue solid lines correspond to the present numerical prediction computed through~\eqref{eq:WNLeps3_stab_bound} for $Re=10^6$, while the red dash-dot lines denote the present Faraday tongues shifted by 5\%. Black and orange dash-dot lines as well as blue solid lines have been added on top of the original figure from K13.}
    \label{fig:tongues_teta90_Kidambi13}
\end{figure}

The most comprehensive investigation of Faraday thresholds with pinned contact line the authors are aware of is that of K13 (see table~\ref{tab:literature_tab}), who considered the case of a perfect brimful condition (meniscus--free) in the inviscid limit. Unlike the classic case of an ideal moving contact line, K13 showed that the pinned contact line problem can be recast into an infinite system of coupled Mathieu equations taking the following form
\begin{equation}
\label{eq:Kidambi_Mathieu}
\frac{d^2\mathbf{y}}{d\tau^2}+\left(\mathbf{P}-2\mathbf{Q}\cos{2\tau}\right)\mathbf{y}=\mathbf{0},
\end{equation} 
\noindent where matrices $\mathbf{P}$ and $\mathbf{Q}$, obtained via projection onto the test function space, are in general not diagonal (for a free contact line $\mathbf{P}$ and $\mathbf{Q}$ are diagonal, so that~\eqref{eq:Kidambi_Mathieu} reduces to~(2.14) of \cite{Benjamin54}, i.e. uncoupled Mathieu equations). Three different methods \citep{nayfeh1995nonlinear} can be used to solve~\eqref{eq:Kidambi_Mathieu} , namely, (i) the mapping at a period (given by the Floquet theory), (ii) the Hill's infinite determinant method (used by \cite{Kumar1994}) and (iii) the multiple scale method. The first two techniques were used in K13 and, particularly, the first one was employed in order to describe the so-called combination resonance tongues (indicated by the black arrows in figure~\ref{fig:tongues_teta90_Kidambi13}), which are not studied in the present work focused on sub--harmonic tongues only (see K13 for a thorough discussion). The disadvantage of the multiple scale method is that generally it is not suitable for the exploration of a large part of the parameter space, however, as anticipated in the introduction, the application of the first two techniques is challenging when the initial free surface is not assumed to be flat. Here we use the inviscid results provided by K13 to validate the present WNL model for prediction of sub--harmonic instability onset in the limit of high Reynolds numbers (e.g. $Re$ is assumed to be $\sim10^6$ in the present viscous analysis). A quantitative comparison of the prediction of sub--harmonic Faraday thresholds with results by K13 is shown in figure~\ref{fig:tongues_teta90_Kidambi13} for $\theta_s=90^{\circ}$, $h/R=H=1$, for two different Bond numbers and for two non-axisymmetric modes, i.e. $\left(1,1\right)$ and $\left(1,2\right)$. For computational reasons, the instability regions (gray shaded) computed by K13 were obtained by truncating the number of basis function $N_{K13}$ to 2, although convergence of the natural frequencies was achieved by taking $N_{K13}=30$, as stated by K13 in his table~1, causing a systematic underestimation of approximately 5\%. The vertical black dash-dot lines, corresponding to the converged natural frequencies reported in table~1 for $N_{K13}=30$, agrees perfectly with the present prediction, which prescribes the correct slope of the right and left marginal stability boundaries (blue solid lines). If the present prediction is shifted by -5\% (orange dash-dot lines), results match. We can hence conclude that the present model is congruent with the analysis by K13 and it prescribes correctly the sub--harmonic Faraday tongues for a pinned contact line case in the limit of validity of the WNL model, i.e. small external forcing amplitude and small detuning.

\subsubsection{Brimful condition: comparison with recent experiments by S21 for $\theta_s=90^{\circ}$}\label{subsubsec:Sec5subsec6subsubsec1}

\begin{figure}
    \centering
    \includegraphics[width=1.\textwidth]{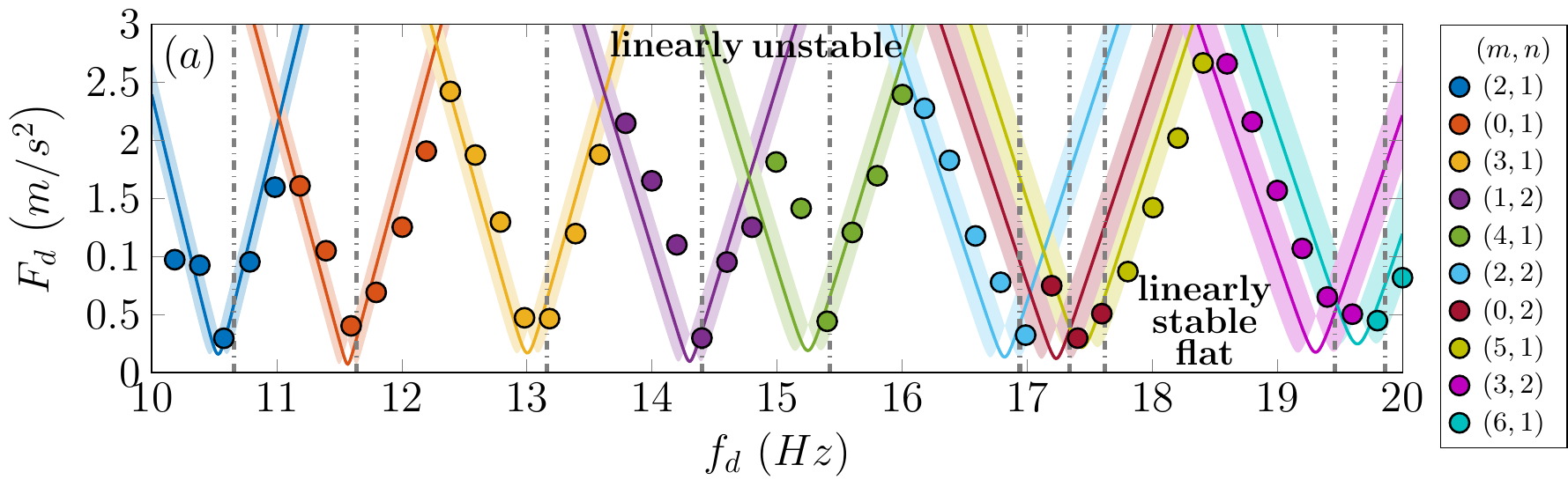}
    \includegraphics[width=1.\textwidth]{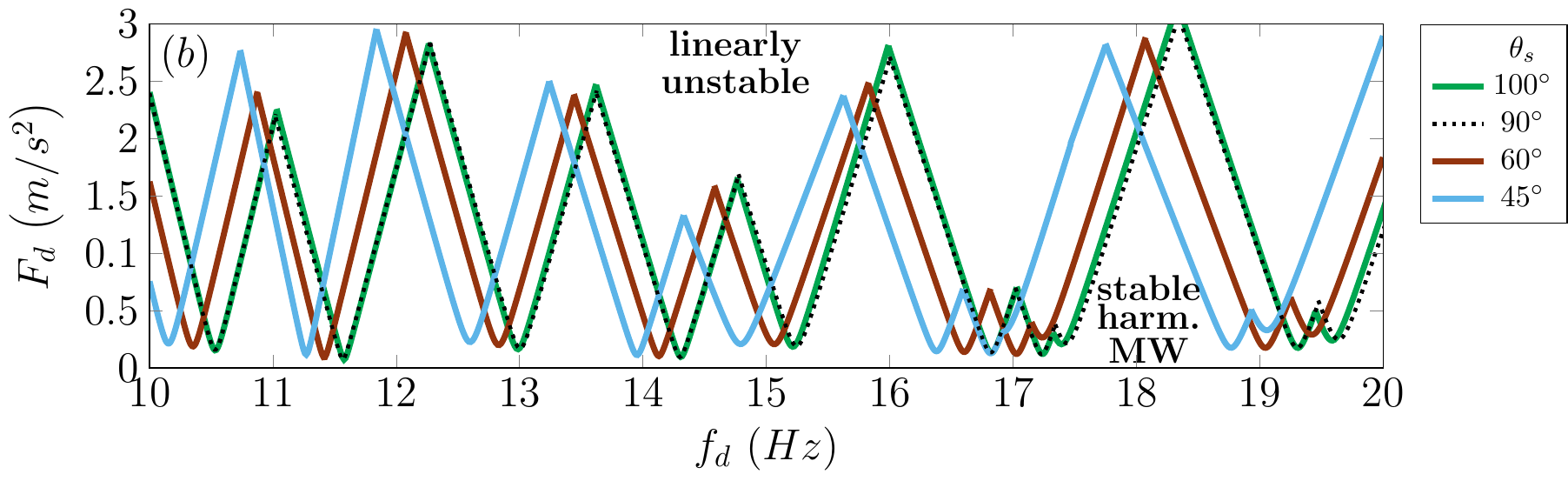}
    \caption{(a) Colored solid lines: boundaries of the sub--harmonic Faraday tongues predicted by~\eqref{eq:WNLeps3_stab_bound} in the forcing acceleration amplitude-forcing frequency dimensional space, $\left(f_d,F_d\right)$. Here the static contact angle was set to $\theta_s=90^{\circ}$. The colored filled circles corresponds to the original experimental values extracted from figure~4 of S21 and corresponding to different waves $\left(m,n\right)$. The black dash-dot lines correspond to their inviscid numerical calculation. Parameters: $R=0.035\,\text{m}$, $h=0.022\,\text{m}$, $\rho=1000\,\text{kgm$^{-3}$}$, $\mu=0.001\,\text{kgm$^{-1}$s$^{-1}$}$ and $\gamma=0.072\,\text{Nm$^{-1}$}$, for which $Bo=166.2$ and $Re=20\,437$. Colored bands: marginal stability boundaries computed for a container radius $R=\left(0.035-0.0003\right)\,\text{m}$ (right boundary) and $R=\left(0.035+0.0003\right)\,\text{m}$ (left boundary). (b) Modification of the linearly unstable regions due to contact angle effects, where the results for three values of $\theta_s$, including $90^{\circ}$ (black dotted lines) as in (a), are compared for a nominal radius $R=0.035\,\text{m}$.}
    \label{fig:tongues_teta90}
\end{figure}

\noindent From the knowledge of the authors, no systematic and formal calculation of the linear sub--harmonic Faraday tongues for pinned contact line and including viscous dissipation are available in literature, even for the simpler case of a flat static free surface, $\theta_s=90^{\circ}$. With regard to small circular--cylinder experiments, this configuration was recently studied by \cite{shao2021surface} (S21). By properly filling the container they could reproduce an initially flat static free surface, which remains stable and flat below Faraday threshold and thereby they could derive experimentally the boundaries of the unstable regions. Their experimental measurements (extracted from figure~4 of S21) are illustrated in figure~\ref{fig:tongues_teta90}(a), as colored filled circle, together with our numerical prediction from~\eqref{eq:WNLeps3_stab_bound} (colored solid lines). \cite{shao2021surface} also employed a Rayleigh--Ritz approach \citep{bostwick2009capillary} to estimate numerically the natural frequency in the inviscid limit and this result, which showed a good agreement with their experiments, is reported for completeness in figure~\ref{fig:tongues_teta90}(a) as vertical black dash-dot lines.\\
\indent Qualitatively speaking, our numerical analysis for $\theta_s=90^{\circ}$ correctly predicts the occurrence of the same sub--harmonic single-mode instabilities in the selected frequency window. In agreement with experimental observations, the viscous WNL analysis prescribes an onset acceleration nearly constant for all $\left(m,n\right)$-modes in the range $f_d\in\left[10,20\right]$ Hz with a discrete spectrum of sub--harmonic resonances. Such a well-defined quantization is seen to persist even for higher frequency. Indeed, despite the occurrence of mode competitions, detection of harmonic responses and overlap of instabilities due to cluster of tongues, e.g. for $f_d\in\left[16.5\,\text{Hz},17.5\text{Hz}\right]$, such a peculiar feature allowed \cite{shao2021surface} to experimentally observe up to 50 different modes in the frequency-band ($7-47\,\text{Hz}$). However, they did not report detailed measurements of the Faraday tongues for $f_d<10$ and $>20$ Hz. In Appendix~\ref{sec:appB}, using equation~\eqref{eq:WNLeps3_stab_bound}, we tentatively reconstructed the whole sub--harmonic spectrum investigated by \cite{shao2021surface}. The present numerical prediction is seen to be fully consistent with their observations in the entire range of frequency.\\
\indent From a quantitative perspective, all the experimental frequencies are slightly larger than the ones predicted here and this shift is roughly of the order of +1\% for all measurements. We note that, since viscous dissipation for this case is very small, the location of a sub--harmonic minimum threshold essentially depends on the natural frequency only, which remain very close to their inviscid approximations. In \S\ref{sec:Sec4} we compared our results with several previous experiments and, while discrepancies were observed in terms of damping between different approaches, the prediction of natural frequencies was generally within 0.6-0.7\% of the experimental values and in excellent agreement with previous theoretical predictions. The inviscid calculation proposed by \cite{shao2021surface} matches well the experimental resonance frequencies (see black dash-dot lines in figure~\ref{fig:tongues_teta90}(a)), although their calculation seems to produce slightly larger values when compared with the present paper and with previous studies, e.g. experimental and inviscid predictions by \cite{henderson1994surface}. It is difficult to attribute a positive +1\% shift to a specific cause, especially because the pinned contact line configuration is known to produce the largest frequencies among the possible contact line boundary conditions, e.g. a free contact line. Presence of free surface contamination (surface film) is expected to  

\clearpage
\begin{table}
\centering
\begin{tabular}{c|cccc|ccc|c}
$\left(m,n\right)$ & & $\lambda^{90^{\circ}}$ & $\zeta^{90^{\circ}}$ & & $\lambda^{45^{\circ}}$ & $\zeta^{45^{\circ}}$ & & $\omega_{R=0.0347\,\text{m}}^{90^{\circ}}$ \\ \hline
$\left(2,1\right)$ & & -0.0066+$\text{i}\,$1.9773 & -0.0046-$\text{i}\,$0.4094 & & -0.0085+$\text{i}\,$1.9063 & -0.0059-$\text{i}\,$0.3889 & & 1.9816\\
$\left(0,1\right)$ & & -0.0034+$\text{i}\,$2.1716 & -0.0032-$\text{i}\,$0.4594 & & -0.0042+$\text{i}\,$2.1158 & -0.1255-$\text{i}\,$0.3319 & & 2.1752\\
$\left(3,1\right)$ & & -0.0082+$\text{i}\,$2.4414 & -0.0068-$\text{i}\,$0.4739 & & -0.0109+$\text{i}\,$2.3653 & -0.0096-$\text{i}\,$0.4744 & & 2.4466\\
$\left(1,2\right)$ & & -0.0051+$\text{i}\,$2.6849 & -0.0028-$\text{i}\,$0.5210 & & -0.0060+$\text{i}\,$2.6194 & -0.0057-$\text{i}\,$0.5234 & & 2.6903\\
$\left(4,1\right)$ & & -0.0099+$\text{i}\,$2.8624 & -0.0084-$\text{i}\,$0.5166 & & -0.0138+$\text{i}\,$2.7771 & -0.0176-$\text{i}\,$0.6464 & & 2.8695\\
$\left(2,2\right)$ & & -0.0076+$\text{i}\,$3.1577 & -0.0026-$\text{i}\,$0.5562 & & -0.0087+$\text{i}\,$3.0753 & -0.0064-$\text{i}\,$0.5865 & & 3.1659\\
$\left(0,2\right)$ & & -0.0069+$\text{i}\,$3.2352 & -0.0025-$\text{i}\,$0.5657 & & -0.0079+$\text{i}\,$3.1578 & -0.0076-$\text{i}\,$0.6095 & & 3.2436\\
$\left(5,1\right)$ & & -0.0120+$\text{i}\,$3.2734 & -0.0095-$\text{i}\,$0.5461 & & -0.0169+$\text{i}\,$3.1766 & -0.0138-$\text{i}\,$0.5239 & & 3.2828\\
$\left(3,2\right)$ & & -0.0103+$\text{i}\,$3.6245 & -0.0029-$\text{i}\,$0.5777 & & -0.0119+$\text{i}\,$3.5236 & -0.0023-$\text{i}\,$0.6632 & & 3.6360\\
$\left(6,1\right)$ & & -0.0143+$\text{i}\,$3.6872 & -0.0103-$\text{i}\,$0.5665 & & -0.0202+$\text{i}\,$3.5780 & -0.0148-$\text{i}\,$0.6026 & & 3.6995\\
\end{tabular}
\caption{Nondimensional natural frequencies, damping coefficients ($\lambda$ is the eigenvalue $\lambda=-\sigma+\text{i}\omega$) and complex normal form coefficient $\zeta=\zeta_{\text{R}}+\text{i}\zeta_{\text{I}}$ for both $\theta_s=90^{\circ}$ and $\theta_s=45^{\circ}$, associated with the modes shown in figure~\ref{fig:tongues_teta90} and computed for $R=0.035\,\text{m}$, $h=0.022\,\text{m}$, $\rho=1000\,\text{kgm$^{-3}$}$, $\mu=0.001\,\text{kgm$^{-1}$s$^{-1}$}$ and $\gamma=0.072\,\text{Nm$^{-1}$}$, for which $Bo=166.2$ and $Re=20\,437$. In the last column the natural frequency computed for $\theta_s=90^{\circ}$ and $R=0.035-0.0003=0.0347\,\text{m}$ (associated to the colored solid lines in figure~\ref{fig:tongues_teta90}(a)) are reported. The number of points in the radial and axial directions for the GLC grid used is this calculation is $N_r=N_z=80$, for which convergence is achieved.}
\label{tab:conv_amp_eq_coeff_fig6}
\end{table}

\clearpage
\noindent slightly increase the rigidity of the free surface, leading to higher resonance frequencies, but also to larger damping coefficients \citep{Miles67,henderson1990single,henderson1994surface}. However, the authors do not report any evidences of surface contamination. In the present case, such a slight systematic mismatch is more likely to be caused by little incongruities between numerics and experiments. For instance, in this case the Bond number is relatively low, $Bo=166.2$, so that little variations in the value of the surface tension or geometrical tolerances on the container radius and depth moving from experiments to numerics could contribute to shift the tongues slightly. We tentatively attributed such a mismatch to a geometric tolerance on the container radius, $R$. In fact, a tolerance of $\pm0.0003\,\text{m}$, i.e. the numerical radius is set to $R=0.0347\,\text{m}\approx 0.035\,\text{m}$, is sufficient to produce a +1\% frequency shift (see colored bands in figure~\ref{fig:tongues_teta90}(a), where the right boundaries correspond to $-0.0003\,\text{m}$ (+1\%), whereas the left ones to $+0.0003\,\text{m}$ (-1\%)) and the agreement is remarkably good.\\
\indent Above all, we see that the WNL model predicts correctly the coefficient $\zeta$, which prescribes the slope of the transition curves for all tongues. For completeness, the value of the damping coefficients, natural frequencies and of the normal form coefficient $\zeta$ for two different static contact angles used in figure~\ref{fig:tongues_teta90} are given in table~\ref{tab:conv_amp_eq_coeff_fig6}, where the natural frequencies computed for the container radius $R=0.0347\,\text{m}$ (figure~\ref{fig:tongues_teta90}(a)) is also given (last column).

\subsubsection{Nearly--brimful condition: static contact angle effects and meniscus waves modifications}\label{subsubsec:Sec5subsec6subsubsec2}

When the value of the prescribed static contact angle is $\theta_s\ne90^{\circ}$, then the initial static free surface is not flat, but rather concave ($\theta_s<90^{\circ}$) or convex ($\theta_s>90^{\circ}$), and its effects on Faraday waves can be studied exploiting the present WNL analysis. In \S\ref{sec:Sec4} we discussed how the static meniscus modifies the natural frequencies and damping coefficients in a non trivial way depending on the wavenumber of the mode, on the Bond and Reynolds number and on the fluid depth \citep{kidambi2009meniscus} (K09). Moreover, under vertical oscillations, the meniscus emits axisymmetric traveling waves (see figure~\ref{fig:eps1_eps2_sol_surf_mn32_mn02}(c), (i) and (p)), which, with the WNL scaling adopted in this work, are coupled at third order with the sub--harmonic parametric waves and hence contribute to alter the instability regions. With regard to the same configuration of figure~\ref{fig:tongues_teta90}(a) \citep{shao2021surface}, in figure~\ref{fig:tongues_teta90}(b) we examine the influence of these capillary effects on the linear Faraday thresholds. For this configuration the natural frequencies are found to have a maximum for $\theta_s\approx90^{\circ}$ (similarly to figure~\ref{fig:expPicard}(b) \citep{picard2007resonance} (PD07). This suggests that the little shift (+1\%) in the experimental measurements reported in figure~\ref{fig:tongues_teta90}(a) is not due to an uncontrolled nearly--brimful condition. When the static contact angle $\theta_s$ is decreased the meniscus introduces a negative shift in all Faraday tongues, which also show a slightly higher onset acceleration owing to an increase of the dissipation occurring in the meniscus region (in spite of the fact that the frequency is lower). For $\theta_s>90^{\circ}$, e.g. $100^{\circ}$, the onset is slightly lowered (slight decrease of the dissipation occurring in the meniscus region, in agreement with experimental observation by \cite{henderson1992effects}). As a result of the mode shape modification by contact angle effects (see figure~\ref{fig:eps1_eps2_sol_surf_mn32_mn02}(a), (g) and (n)) and of the third order coupling with harmonic meniscus waves, the slope of the transition curves is also altered, but only slightly. In other words, harmonic meniscus waves do not affect significantly the linear instability onsets of these sub--harmonic resonances. This observation is in agreement with \cite{batson2013faraday}, who noticed that a significant meniscus modification is more likely to occur for harmonic Faraday waves and particularly for axisymmetric $\left(0,n\right)$ modes. This is somewhat intuitive as meniscus waves, being axisymmetric and having zero threshold, are essentially indistinguishable from harmonic axisymmetric parametric waves when the driving angular frequency is $\Omega_d=\omega_{0n}$. Notwithstanding that the coupling between meniscus and sub--harmonic-parametric waves is only weak, the shift in frequency may lead to a reorganization of the discrete spectrum. This is observable in figure~\ref{fig:tongues_teta90}(b) for modes $\left(0,2\right)$ and $\left(5,1\right)$. Decreasing $\theta_s$, the region associated with mode $\left(5,1\right)$ progressively lies within that of mode $\left(0,2\right)$ and possibly disappears. Having a higher onset acceleration, is less likely to be detected. This reorganization is expected to be more pronounced for higher frequency modes, where, for a fixed Bond number, the characteristic mode wavelength becomes comparable and eventually smaller than the characteristic meniscus length, i.e. the (static) capillary length $l_c\sim1/\sqrt{Bo}$, thus enhancing contact angle effects. Lastly, it should be noted that that although parametric waves are linearly stable for all $\theta_s$ outside the Faraday tongues, the free surface (which is maintained flat when $\theta_s=90^{\circ}$) appears as the superimposition of the static meniscus and harmonic meniscus waves, whose amplitude (for a fixed frequency) is proportional to the forcing amplitude, giving rise to an imperfect bifurcation diagram that shows a tailing effect and that will be examined in the following.

\subsection{Weakly nonlinear threshold and bifurcation diagram}\label{subsec:Sec5subsec8}

\begin{figure}
    \centering
    \includegraphics[width=1\textwidth]{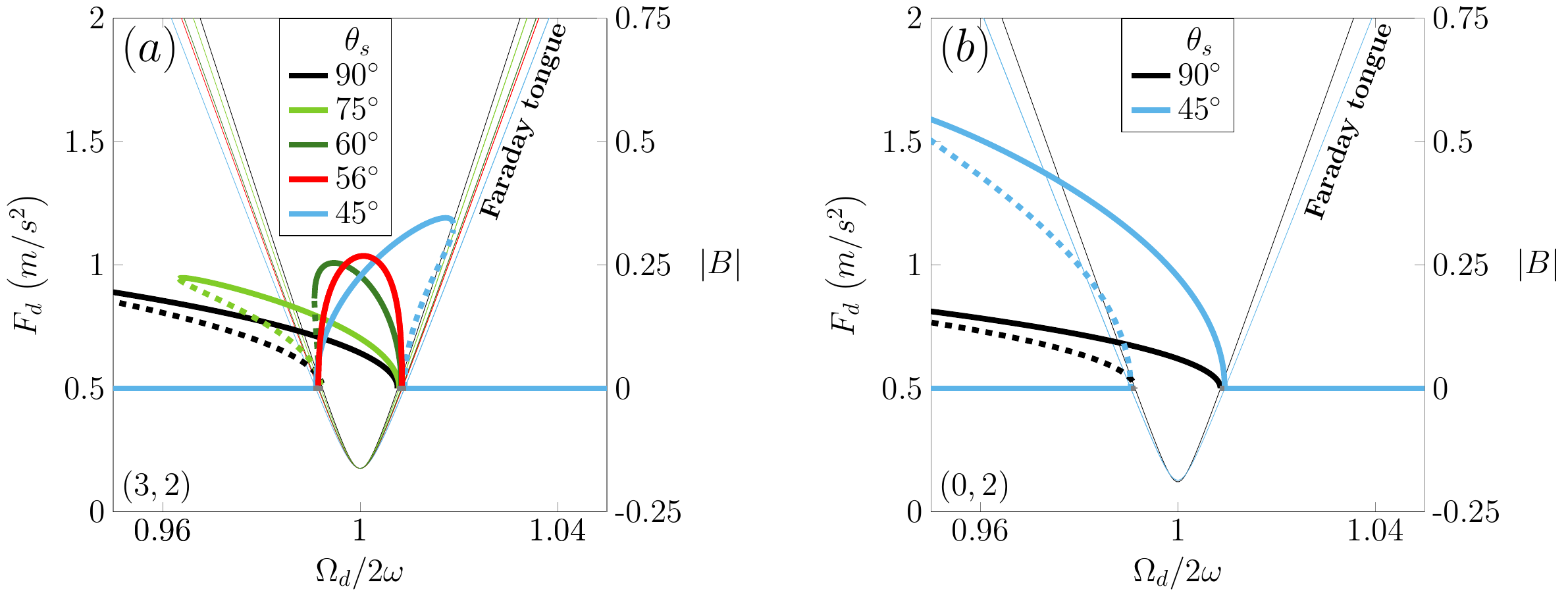}
     \caption{Linear acceleration threshold (Faraday tongue) (left-y-axis, thin solid lines) and saturated wave amplitude, $|B|$, (right-y-axis, thick solid lines) for a fixed acceleration amplitude $F_d=0.5\,\text{ms$^{-2}$}$, while the driving frequency is varied. Stable branches for $|B|$ are shown as solid lines, while unstable branches as dashed lines. Two different modes corresponding, namely (a) $\left(m,n\right)=\left(3,2\right)$ and (b) $\left(0,2\right)$, are shown. Different static contact angle are considered. The frequency is normalized with twice the natural frequency of the corresponding excited mode, so that the lowest linear threshold occurs for $\Omega/2\omega=1$ for all $\theta_s$. At convergence (GLC grid $N_r=N_z=80$), the complex nonlinear amplitude equation coefficient, $\chi=\chi_R+\text{i}\,\chi_I$, for mode $\left(0,2\right)$ (subplot (b)), assumed the values, $\chi^{90^{\circ}}=-0.0909-\text{i}\,1.9094$ and $\chi^{45^{\circ}}=-0.0184-\text{i}\,0.5617$. Geometrical and physical parameters are set as in figure~\ref{fig:eps1_eps2_sol_surf_mn32_mn02}.}
    \label{fig:bifurcation_diag_WNL1}
\end{figure}

In this paragraph, we focus on the stability of the non-trivial equilibrium, $|B|\ne0$, of system~\eqref{eq:WNLeps3_ampB}-\eqref{eq:WNLeps3_phaseB}. Again, for stationary solutions, we find by eliminating $\Theta$ that
\begin{equation}
\label{eq:WNLeps3_sol_absB}
c_3|B|^2=-c_1\cos{\left(\varphi_1-\varphi_3\right)}\pm\sqrt{c_2^2F^2-c_1^2\sin^2{\left(\varphi_1-\varphi_3\right)}},
\end{equation}
\noindent with physical real solutions for $F\ge\frac{c_1}{c_2}|\sin{\left(\varphi_1-\varphi_3\right)}|$. This well-known result prescribes either a supercritical or a subcritical transition when the marginal stability boundaries are crossed, i.e. by changing forcing frequency and amplitude. The location of the hysteresis depends on the sign of the nonlinear coefficient $\chi$ \citep{Kovacic2018}, which assumes the meaning of a nonlinear detuning, while the boundary of the hysteresis region in the parameter space is defined by the nonlinear threshold 
\begin{equation}
\label{eq:WNL_th_non}
F_{th}^{NL}=c_1/c_2|\sin{\left(\varphi_1-\varphi_2\right)}|,
\end{equation}
\noindent In figure~\ref{fig:bifurcation_diag_WNL1} the nonlinear wave amplitude saturation, for a fixed external acceleration amplitude, $F_d$, and for a varying excitation frequency, $\Omega_d$, is shown for two different modes, $\left(3,2\right)$ and $\left(0,2\right)$, and for different static contact angle values. The linear acceleration threshold (Faraday tongue) is plotted versus a normalized driving frequency in order to better compare the difference between the two cases with $\theta_s=90^{\circ}$ (flat static surface, brimful condition) and $45^{\circ}$ (static meniscus and meniscus waves, nearly--brimful condition). As previously discussed, contact angle modifications on the linear thresholds are only weak. When a concave ($\theta_s<90^{\circ}$) static meniscus is considered, the damping is generally higher, the shape of the mode is, however, modified, leading to a slightly different value of the complex linear coefficient $\zeta$ (see table~\ref{tab:conv_amp_eq_coeff_fig6}), which also englobes the second order coupling between parametric and meniscus waves. As a consequence, the minimum onset acceleration, given by the ratio $\sigma/|\zeta|$, is often comparable.\\
\indent Supercritical and subcritical bifurcations of Faraday waves have been widely discussed in literature (see for instance \cite{Douady90,Rajchenbach2015} among other references), hence we limit here to recall that if $\cos{\left(\varphi_1-\varphi_3\right)}>0$, or alternatively $\Lambda=\Omega_d-2\omega>-2\sigma \chi_R/\chi_I$, then the bifurcation is supercritical, while if $\cos{\left(\varphi_1-\varphi_3\right)}<0$, or $\Lambda=\Omega_d-2\omega<-2\sigma \chi_R/\chi_I$, the transition is subcritical, the sign of $\chi_I$ determines whether hysteresis occurs on the left-side or on the right-side. The inferior boundary of the hysteresis region in the $\left(\Omega_d,F_d\right)$-plane is defined by equation~\eqref{eq:WNL_th_non}. In other words, the ratio $\chi_R/\chi_I$, through the relation $\varphi_3=\tan^{-1}{\left(\chi_I/\chi_R\right)}$, determines the importance of the subcritical region in the parameter space \citep{Douady90,hsu1977nonlinear,nayfeh1995nonlinear,Meron87,gu1987resonant,Douady90}.\\
\indent We underly that the amplitude equation coefficients setting the nonlinear threshold and the bifurcation diagram are not calibrated from experimental data, but their values are here computed numerically from first principles through our WNL analysis.

\subsubsection{Wave amplitude increase and sub--criticality suppression}\label{subsec:Sec5subsec8subsubsec1}

We now discuss contact angle modifications on the nonlinear wave amplitude saturation in comparison with the results for the classic case with $\theta_s=90^{\circ}$ (flat static interface). A first striking result is shown in figure~\ref{fig:bifurcation_diag_WNL1}(b) for the second axisymmetric mode $\left(0,2\right)$, which displays the bifurcation diagram (in the right y-axis) computed by sweeping the external forcing frequency at a fixed forcing amplitude, i.e. $F_d=0.5\,\text{m/s$^{2}$}$ (left y-axis). Figure~\ref{fig:bifurcation_diag_WNL1}(b) shows that, despite contact angle effects do not alter substantially the sub--harmonic Faraday tongue (the unstable region is slightly wider), presence of the meniscus waves, from which the parametric wave bifurcates, can strongly increase the wave amplitude response (up to three times in this case). The magnitude of such an increase is found to be maximum for axisymmetric waves. Again, this can be intuitively explained by considering that axisymmetric parametric and meniscus waves share the same spatial symmetries, despite their different nature, i.e. sub--harmonic versus harmonic responses. Therefore, axisymmetric parametric waves, which emerge on the top of meniscus waves, appear to be nonlinearly more destabilized by the latter when compared to other modes.\\
\indent The second interesting result is shown in figure~\ref{fig:bifurcation_diag_WNL1}(a). In some cases, as for example for mode $\left(3,2\right)$, we observe an inversion of the bifurcation diagram, caused by the change of sign of the nonlinear coefficient, $\chi$, as the static contact angle is varied from $90^{\circ}$ to $45^{\circ}$ (same extrema of figure~\ref{fig:bifurcation_diag_WNL1}(b)). This is mathematically not paradoxical as one more independent parameter, i.e. the contact angle $\theta_s$, is added to the overall parameter space. The increase of the wave amplitude response with a decrease of $\theta_s$ is accompanied by a progressive reduction of the region of hysteresis, until a threshold value, $\theta_s^{th}$ ($=56$ for the case of figure~\ref{fig:bifurcation_diag_WNL1}(a)), is reached. Eventually, the direction of the bifurcation reverses and the size of the hysteresis region starts to increase again. At the threshold value, $\theta_s^{th}$, corresponding to figure~\ref{fig:bifurcation_diag_WNL1}(a), the nonlinear coefficient $\chi$ takes the value $\chi=-0.0729+\text{i}\,0.0083$, yielding a large ratio $\chi_R/\chi_I$ in absolute value), for which the phase $\varphi_3$ is nearly $-\pi$, thus meaning that the sub--criticality is totally suppressed and the bifurcation is always supercritical for each combination of external control parameter in $\left(f_d,F_d\right)$-plane \citep{Douady90}. From the knowledge of the authors, such a contact--angle--related behaviour has not been reported in the literature yet, thus suggesting a pursuable direction that future lab--scale and controlled experiments could undertake. 

\subsubsection{The imperfect bifurcation diagram: \textit{tailing }effect}\label{subsec:Sec5subsec8subsubsec2}

\begin{figure}
    \centering
    \includegraphics[width=1\textwidth]{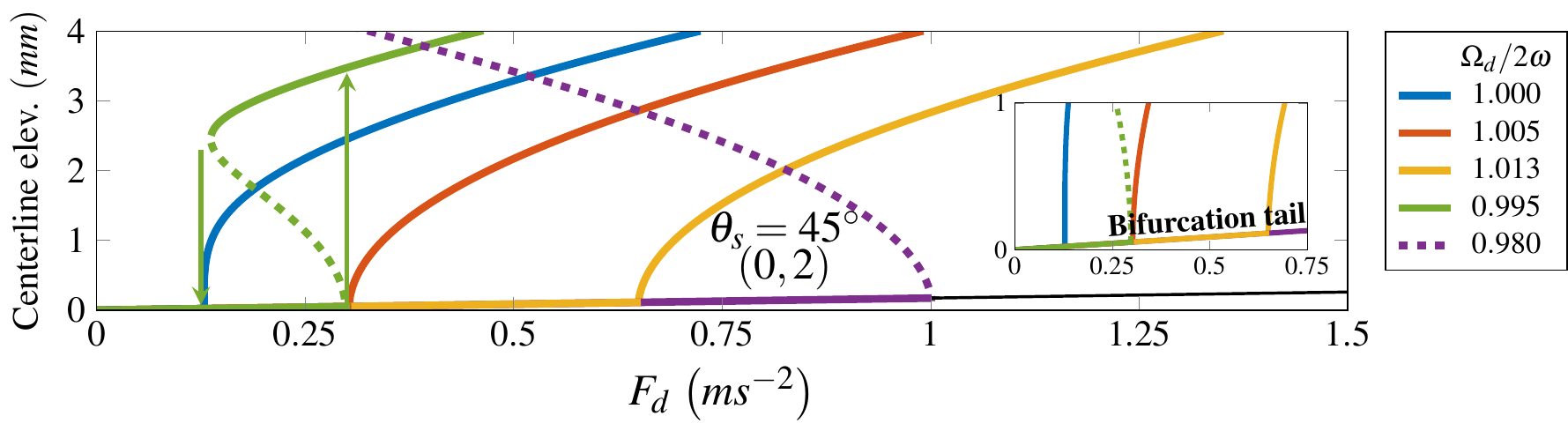}
    \caption{Bifurcation diagram associated with $\left(m,n\right)=\left(0,2\right)$ (see also figure~\ref{fig:bifurcation_diag_WNL1}(b)) and for a static contact angle $\theta_s=45^{\circ}$. Here the total dimensional centerline elevation (axisymmetric dynamic) is reconstructed by summing the various order solutions, i.e. $\eta=\eta_0+\eta_1+\eta_2$ and it is plotted versus the external forcing acceleration for a fixed excitation angular frequency, while different colors correspond to different forcing frequencies. The tailing effect (imperfect bifurcation diagram) produced by presence of harmonic meniscus waves and indicated by the black thin solid line (the amplitude of meniscus waves grows linearly with $F_d$, independently of the parameter combination $\left(F_d,\Omega_d\right)$), is well visible in the right-inset. Colored solid lines are used for stable branches, while colored dashed lines for the unstable ones. The hysteretic loop is indicated by the green arrows.}
    \label{fig:bifurcation_diag_WNL2}
\end{figure}

As shown in figure~\ref{fig:bifurcation_diag_WNL1}, the linear threshold given by \eqref{eq:WNLeps3_stab_bound} prescribes a stable solution outside the sub--harmonic Faraday tongues (see figure~\ref{fig:tongues_teta90}) with a stationary mode amplitude $|B|=0$.  Nevertheless, we remind the reader that the total solution, e.g. in terms of free surface elevation, is given by the sum of the solutions at the various orders in $\epsilon$, i.e. $\eta=\eta_0+\eta_1\left(|B|\right)+\eta_2\left(F_d,B^2,|B|^2\right)$. In particular, meniscus waves, whose amplitude is proportional to the external acceleration amplitude, $F_d$, are contained in the second order response $\eta_2$. If one considers an axisymmetric dynamics, e.g. $\left(0,2\right)$, the centerline elevation is a suitable quantitiy to monitor the free surface stability and thus to depict a comprehensive bifurcation diagram. This is done in figure~\ref{fig:bifurcation_diag_WNL2}, where such a bifurcation diagram for $\left(0,2\right)$ is reported for different excitation angular frequencies in a range which gathers both supercritical and subcritical bifurcations. Figure~\ref{fig:bifurcation_diag_WNL2} clearly shows that, when a nearly--brimful condition is considered, e.g. $\theta_s<90^{\circ}$, the sub--harmonic parametric waves, stable outside the Faraday tongues, do not bifurcate from the rest state (as for $\theta_s=90^{\circ}$), but rather from the meniscus waves solution ($\sim F_d$), oscillating harmonically with the driving frequency. This produces a so-called imperfect bifurcation diagram, which displays a tailing effect (highlighted by the black thin solid line) \citep{virnig1988three}. The bifurcation diagram of figure~\ref{fig:bifurcation_diag_WNL2} is also reminiscent of that presented by \cite{batson2013faraday}, although they focus on harmonic parametric waves. For further comments on the imperfect bifurcation diagram are see also Appendix~\ref{sec:appB}.

\section{Validation with axisymmetric direct numerical simulations}\label{sec:Sec6}

\begin{figure}
    \centering
    \includegraphics[width=1\textwidth]{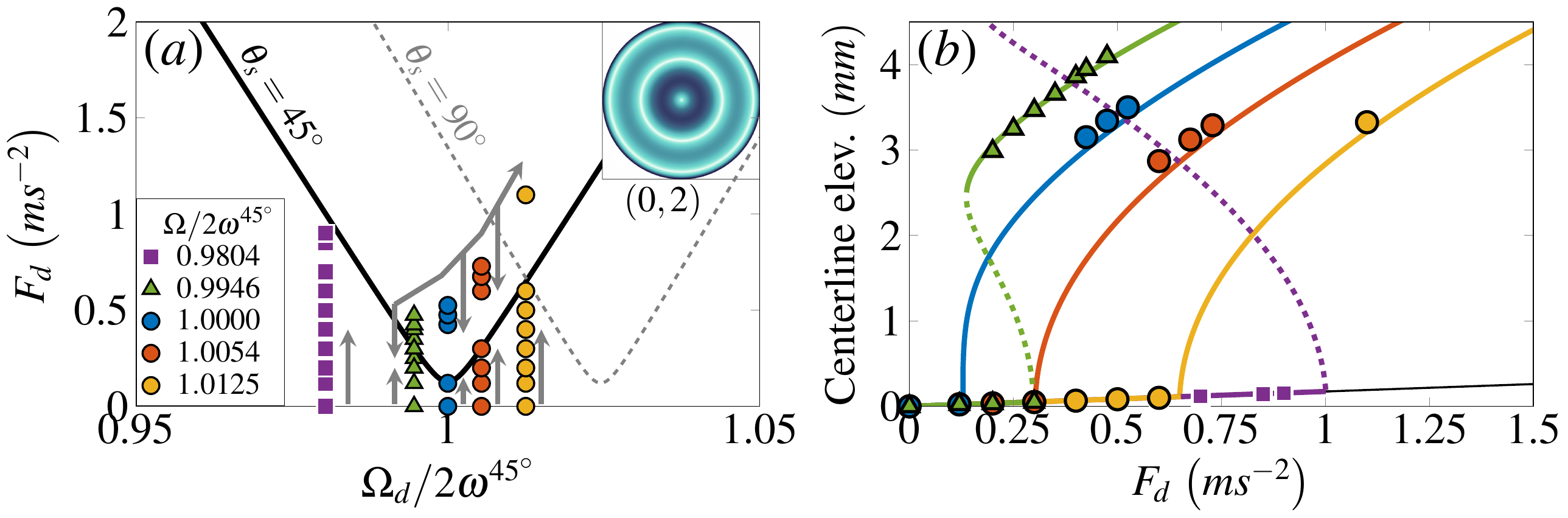}
    \caption{(a) Faraday tongue (black solid line) for the axisymmetric mode $\left(0,2\right)$ and for a static contact angle $\theta_s=45^{\circ}$. Forcing frequency and amplitude in the $\left(f_d,F_d\right)$-space, corresponding to the DNS points in (b), are indicated by colored filled markers. Note that the frequency in the x-axis is normalized using the natural frequency $\omega=3.16$ computed for $\theta_s=45^{\circ}$. The gray arrows denote the direction followed in the continuation procedure for DNS. For completeness, the Faraday tongue for $\theta_s=90^{\circ}$ is reported as gray dashed line. Top-right inset: shape of the normalized $\left(0,2\right)$-eigensurface. (b) Associated bifurcation digram: WNL prediction (lines) versus DNS (markers). The unstable branch is displayed as colored dashed lines. The black solid line indicating the slop of the meniscus wave response is also given to guide the eyes.}
    \label{fig:bif_diag_WNL_DNS}
\end{figure}

In this section, with the purpose of partially validating the weakly nonlinear analysis, we perform nonlinear direct numerical simulations (DNS) associated with the axisymmetric mode $\left(m,n\right)=\left(0,2\right)$, already discussed in \S\ref{sec:Sec5} (see also Appendix~\ref{sec:appB}). Indeed, differently from non-axisymmetric modes $\left(m,n\right)$ that would require computationally demanding full three-dimensional DNS, axisymmetric $\left(0,n\right)$ modes can be solved through axisymmetric DNS, thus reducing the computational burden. To this end, the built-in package for laminar flow with moving interface and automatic remeshing implemented in the finite-element software COMSOL Multiphysics v5.6. were employed. In the underlying problem, we adopted an hybrid quadrilateral-tringular mesh. Specifically, triangular elements were used in the interior, where little deformations occur, while quadrilateral elements were adopted in the neighborhood of the free surface (larger mesh deformation), sidewalls and bottom, where, in addition, boundary layer refinements were used to properly account for the viscous dissipation taking place in the oscillating Stokes boundary layers (see also figure~\ref{fig:numKidambi}). Globally, the grid is made of approximatively $60\,000$ mesh elements. $P_1$--$P_1$ elements (default), stabilized with a streamline diffusion scheme (SUPG, Streamline Upwind Petrov-Galerkin), were used, leading to roughly $230\, 000$ degrees of freedom, for which convergence was tested. Time integration is handled with a mixed-order backward differentiation formula (BDF1/BDF2) with adaptive time-step and the system at each time-step is solved via robust direct method MUMPS (MUltifrontal Massively Parallel sparse direct Solver) coupled with an inner iterative Newton solver.\\
\indent By simulating an axisymmetric dynamics only, all the other non-axisymmetric instabilities are artificially filtered out, i.e. the Faraday tongues for $\left(0,n\right)$ are isolated, enabling a direct comparison of DNS with the single standing-wave expansion adopted in \S\ref{sec:Sec5}. Although such a simplification is not realistic, as often multiple tongues may share nearly the same region of instability and the associated parametric waves may therefore interact nonlinearly, it is extremely convenient for validation purposes and it enables us to easily highlight the various effects, i.e. contact angle and meniscus waves modifications of the Faraday threshold, tackled in in \S\ref{sec:Sec5}.\\

\subsection{Procedure}\label{subsec:Sec6subsec0} 

To start, the shape of the static meniscus, computed in Matlab by solving Eq.~\eqref{eq:static_men} with its boundary conditions (prescribing a static contact angle value, e.g. $\theta_s=45^{\circ}$) was loaded in COMSOL Multiphysics and the static domain was meshed. First simulations were initialized for time $t=0$ with a BDF1 scheme giving a zero velocity field and hydrostatic pressure $p=-z$ as initial conditions. A body forcing, corresponding to the non-dimensional time-dependent gravity acceleration, $-1+\left(F_d/g\right)\cos{\Omega_d t}$, was assigned. The starting point of the gray arrows in figure~\ref{fig:bif_diag_WNL_DNS}(b) indicates the combination of external control parameter $\left(\Omega_d,F_d\right)$ (colored markers), chosen to initiate the simulations, as described above. Once the stationary state for these initial DNS was established, a continuation procedure (directions of the arrows), by slightly adjusting the external amplitude acceleration and angular frequency, was adopted in order to speed up the computations for all the other combinations of parameters here considered (see figure~\ref{fig:bif_diag_WNL_DNS}).

\subsection{Amplitude saturation and free surface reconstruction: WNL vs. DNS}\label{subsec:Sec6subsec1}

The WNL prediction~\eqref{eq:WNLeps3_sol_absB} for the finite amplitude saturation is compared with DNS in figure~\ref{fig:bif_diag_WNL_DNS}.
\begin{figure}
    \centering
    \includegraphics[width=1\textwidth]{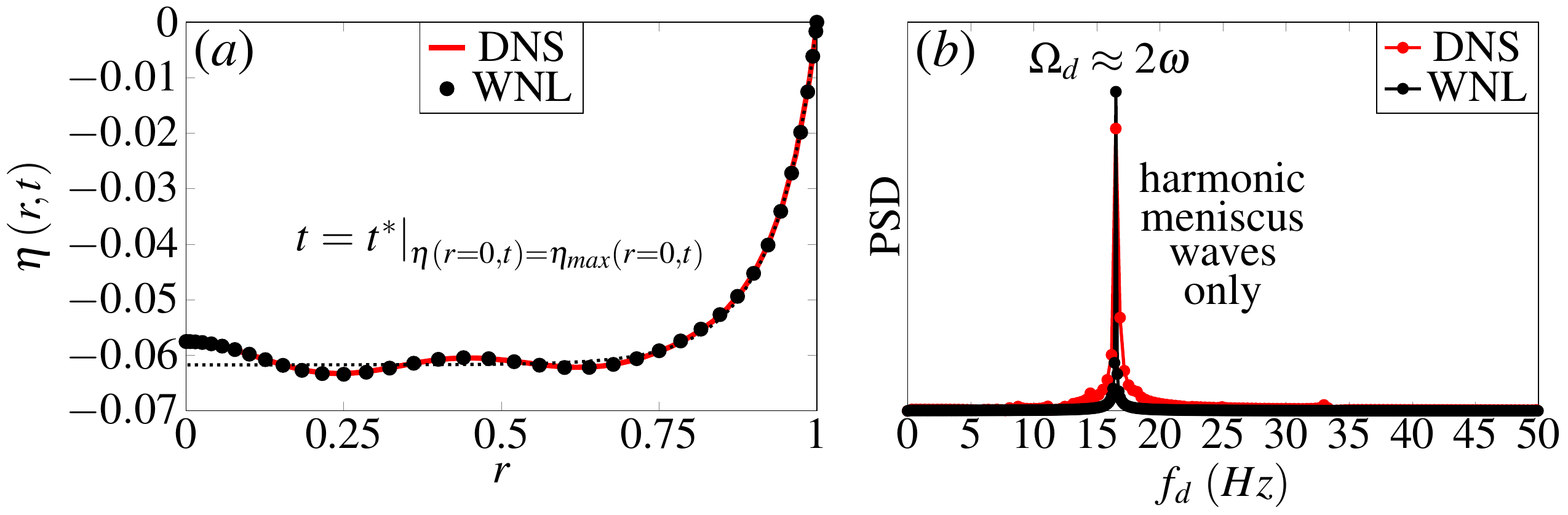}
    \caption{WNL (black) versus DNS (red) below Faraday threshold (outside the Faraday tongue) for $\Omega_d/\omega^{45^{\circ}}=0.9804$ and $F_d=0.85\,\text{ms$^{-2}$}$ (see figure~\ref{fig:bif_diag_WNL_DNS}). (a) Free surface shape computed when the centerline elevation is maximum for. For completeness, the shape of the static meniscus for $\theta_s=45^{\circ}$ is reported as a black dotted line. (b) Corresponding frequency spectrum: power spectral density (PSD) versus dimensional driving frequency, $f_d$.}
    \label{fig:DNS_WNL_men_wave_DYN_FFT}
\end{figure}
\begin{figure}
    \centering
    \includegraphics[width=1\textwidth]{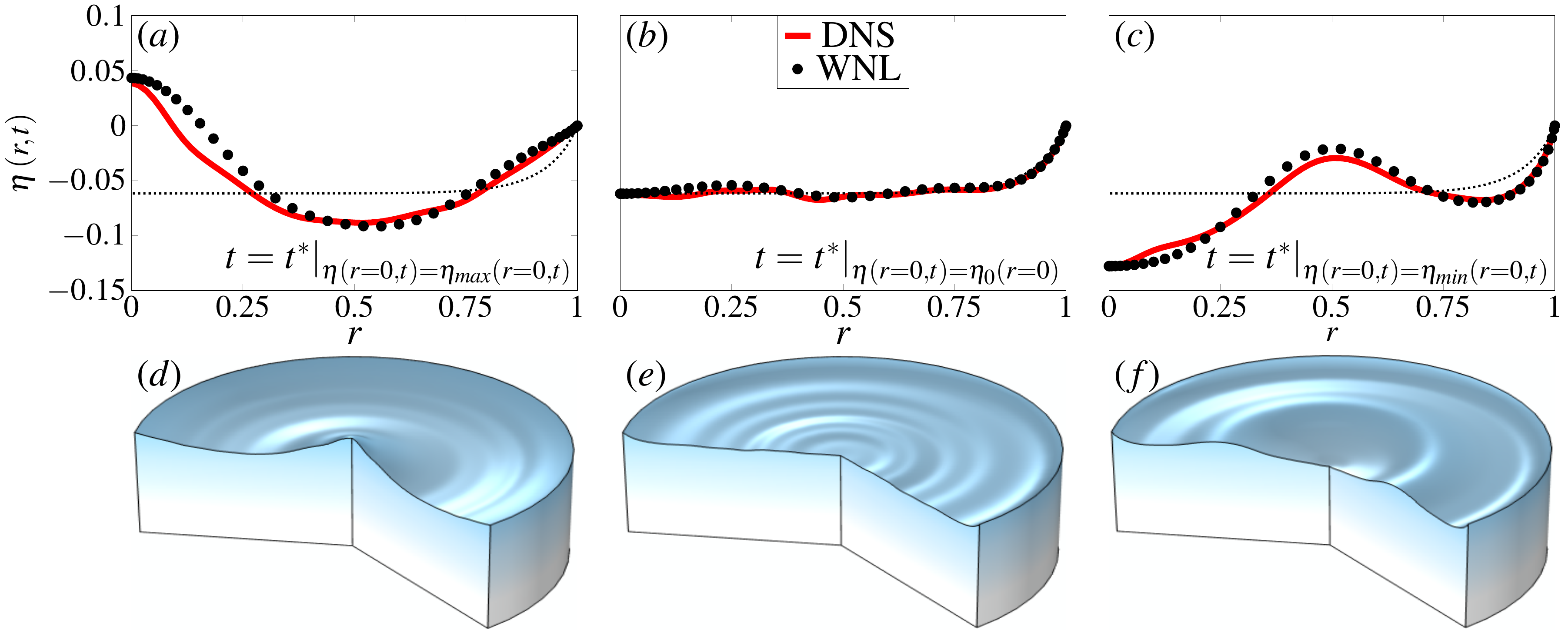}\\
    \medskip
    \includegraphics[width=1\textwidth]{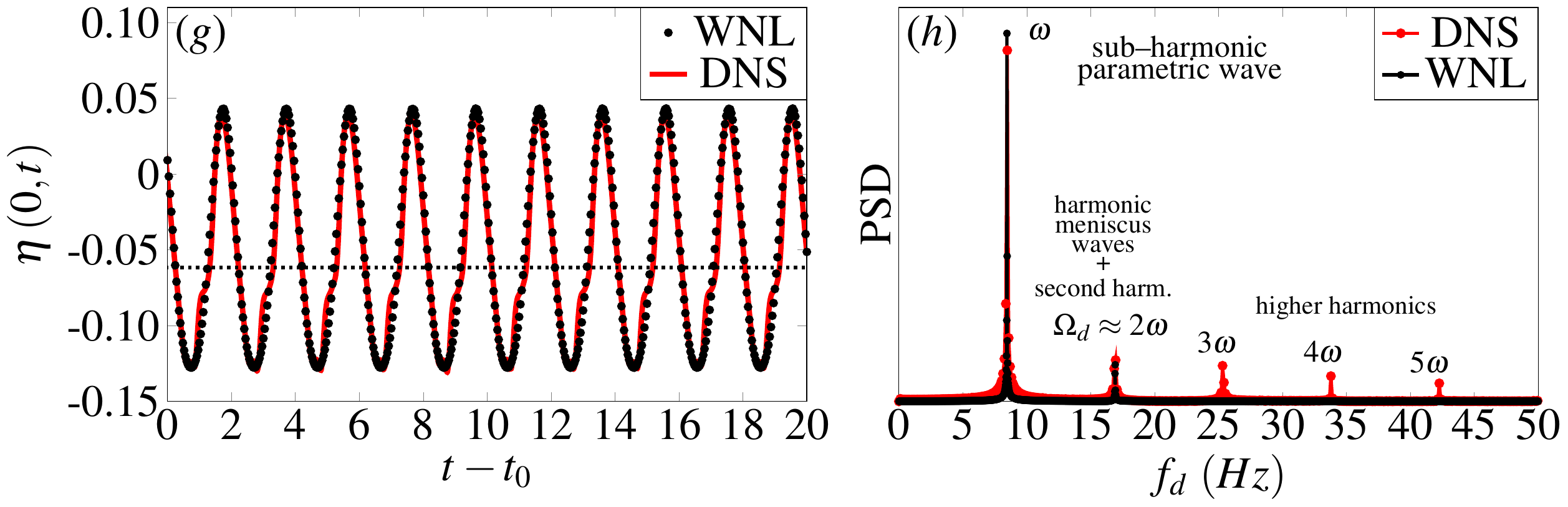}
    \caption{WNL (black) versus DNS (red) above Faraday threshold (within the Faraday tongue) for $\Omega_d/\omega^{45^{\circ}}=1.0054$ and $F_d=0.675\,\text{ms$^{-2}$}$ (see figure~\ref{fig:bif_diag_WNL_DNS}). (a)-(c) Comparison in term of free surface reconstruction for three different time-instants: (a) when the centerline elevation is maximum, (b) when it is zero and equal to the static meniscus position and (c) when it is minimum. For completeness, the shape of the static meniscus for $\theta_s=45^{\circ}$ is reported as a black dotted line. (d)-(f) Full three-dimensional visualization extracted from the DNS. (g) Centerline elevation versus time associated with (a), (b) and (c). $t_0$ is an arbitrary time-instant. The constant value of the static meniscus elevation at $r=0$ is shown as a black dotted line. (h) Frequency spectrum computed from the time-serie shown in (g): power spectral density (PSD) versus dimensional driving frequency, $f_d$.}
    \label{fig:DNS_WNL_param_wave_DYN_3T}
\end{figure}
\noindent The selected combinations of control parameters, i.e. $\left(f_d,F_d\right)$, for DNS calculations are indicated by colored markers in figure~\ref{fig:bif_diag_WNL_DNS}(a), where the gray arrows display the direction followed in the continuation procedure. Once the stationary state is established, i.e. the wave amplitude saturates, and being the underlying dynamics axisymmetric, the centerline free surface elevation is used as reference measure of the free surface destabilization and of its saturation to finite amplitude. The DNS results are therefore compared with the WNL prediction, where the centerline dynamics is reconstructed by evaluating $\eta=\eta_0+\eta_1+\eta_2$ in $r=0$ for any time. The resulting amplitude comparison is shown in figure~\ref{fig:bif_diag_WNL_DNS}(b). At small forcing amplitude below Faraday threshold (see also figure~\ref{fig:bif_diag_WNL_DNS}(b)), only harmonic traveling meniscus waves, whose amplitude is proportional to $F_d$, are observed in the DNS, consistently with the WNL model (straight line in figure~\ref{fig:bif_diag_WNL_DNS}(b)). In this small amplitude regime, the WNL model and DNS fully agree in terms of free surface dynamics, as figures~\ref{fig:DNS_WNL_men_wave_DYN_FFT}(a) and (b) prove. The frequency spectrum in figure~\ref{fig:DNS_WNL_men_wave_DYN_FFT}(b) clearly highlights the harmonic nature of these zero-threhsold meniscus waves, directly forced by the container sidewalls as soon as the vertical excitation starts. 
\indent By increasing the external acceleration amplitude $F_d$, the stability boundary (Faraday tongue in figure~\ref{fig:bif_diag_WNL_DNS}(a)) are eventually crossed and the parametric wave emerge on the top of edge waves, i.e. it bifurcate from the new stable and harmonically oscillating configuration.
\noindent Employing a continuation technique by progressively increasing/decreasing the forcing amplitude at different driving frequencies, several DNS were performed in both the supercritical and subcritical regime (respectively filled colored circles and triangles in figure~\ref{fig:bif_diag_WNL_DNS}). The agreement between DNS and WNL prediction in terms of amplitude saturation is found to be fairly good. Moreover, as figure~\ref{fig:bif_diag_WNL_DNS}(a) shows, DNS are consistent with the frequency shift caused by the presence of the static meniscus for $\theta_s=45^{\circ}$. As an example, the fully nonlinear free surface dynamics obtained from DNS for $\Omega_d/\omega^{45^{\circ}}=1.0054$ and $F_d=0.675\,\text{ms$^{-2}$}$ is compared with the WNL reconstruction in figure~\ref{fig:DNS_WNL_param_wave_DYN_3T}(a)-(c) for three different time-instants, while the corresponding centerline elevation and frequency spectrum are provided in figure~\ref{fig:DNS_WNL_param_wave_DYN_3T}(g) and (h), respectively.
\noindent The WNL model is in agreement with the DNS, which consistently predicts the excitation of a dominant sub--harmonic parametric wave $\left(0,2\right)$, coupled with smaller amplitude harmonic meniscus waves as well as with higher order harmonics (only second harmonics are included in the asymptotic expansion up to the third order in $\epsilon$).\\
\indent As a final comment to this section, while not the purpose of the present analysis, few DNS were performed at higher external acceleration amplitudes, in the parameter region far from the hypotheses of validity of the WNL theory. For the case of figure~\ref{fig:bif_diag_WNL_DNS}(b), preliminary observations revealed that DNS tends to diverge when the centerline elevation approaches a value of approximatively 5 mm, suggesting a potential transition to a highly nonlinear wave-breaking condition and eventually to a finite-time singularity with intense jet formation \citep{basak2021jetting}. See also \cite{das2008parametrically} for a detailed investigation of the occurrence of such a phenomenon in Faraday experiments.


\section{Conclusion}\label{sec:Sec7}

In this paper, we considered sub--harmonic parametric resonances of standing viscous capillary--gravity waves in straight-wall sharp-edged circular--cylindrical containers with brimful (flat static interface) or nearly--brimful (curved meniscus) conditions. First, the numerical tools employed thorough the work were used to compute the natural frequencies and damping coefficients of viscous capillary--gravity waves, which were shown to be in excellent agreement with several experiments, previous theoretical approaches (often based on semi-analytical method involving asymptotic expansions or boundary layer approximations) and numerical models available in the literature. In contradistinction with previous works, the use of numerical scheme based on a full discretization technique allowed us to overcome the mathematical difficulties of formulating an eigenvalue problem for surface waves with a contact line pinned at the brim (which in our case takes the form of a classic generalized linear eigenvalue problem that can be solved numerically with standard techniques) and, at the same time, to easily include in the formulation viscous dissipation and static contact angle effects. This first result opens us to the possibility of investigating several different geometrical configurations of interest, i.e. square cross-sectioned container, as the whole mathematical problem reduces to a meshing problem only.\\
\indent Using the beforehand described tools combined with symbolic calculus, we formalized a numerically-based weakly nonlinear expansion (in the spirit of the multiple timescale method) that provides an amplitude equation for the prediction of sub--harmonic Faraday thresholds of standing waves with pinned--end edge contact line and which corresponds to the classic one widely discussed by \cite{Douady90} and other authors using symmetry arguments solely. However, in this work such amplitude equation has been derived by first principles and the values of the complex normal form coefficients have not a heuristic (or fitting-based) nature, but rather they are obtained in closed form and evaluated numerically.\\
\indent While a simplified version of the underlying fluid problem, i.e. ideal inviscid fluid and perfect brimful conditions ($\theta_s=90^{\circ}$, meniscus--free), was investigated by \cite{Kidambi2013}, in this work we formalized a theoretical and numerical framework that formally accounts for (i) viscous dissipation and (ii) static contact angle effects, including harmonic traveling meniscus waves (nearly--brimful condition), realistic features which are typically encountered in real Faraday experiments. The numerical inviscid analysis by \cite{Kidambi2013} and the recent experimental study by \cite{shao2021surface} were used to validate the WNL model in the simpler case of an initially flat static surface, i.e. no meniscus was present with a static contact angle set to $\theta_s=90^{\circ}$. The agreement with experiments by \cite{shao2021surface} was found to be fairly good in the whole frequency window examined with except for a roughly -1\% frequency-shift with respect to experimental values. Given the excellent agreement, particularly in terms of natural frequencies (typically within 0.6\%), with various experiments discussed in \S\ref{sec:Sec4}, the little shift found in the \textit{vis}-\textit{à}-\text{vis} comparison with \cite{shao2021surface} could not be totally explained and it was tentatively attributed to little geometrical tolerances, e.g. on the container radius. Nevertheless, the slope of the transition curves defining the Faraday tongues as well as the mode dissipation (which determines the lowest linear threshold) were in accordance with experimental measurements. Starting from this reference brimful condition, we progressively introduced in the analysis contact angle effects, simulating the under-filling (or over-filling) of the container. Presence of a static meniscus was shown to determine a negative (at least in the cases examined) frequency shift of all the sub--harmonic Faraday tongues and to slightly increase (or decrease) the minimum onset acceleration, as consequence of a slightly higher (lower) dissipation in the meniscus region, as expected from previous studies. Moreover, sometimes contact angle modifications, modifying the position of the resonances, can induce a reorganization of the frequency spectrum, with some instability lying within other unstable regions, hence making them less likely to be detected.\\
\indent The salient point of the present work is the introduction, within a comprehensive theoretical framework, of harmonic meniscus or edge waves emitted by the oscillating static meniscus under the vertical external excitation, widely discussed in literature, but mostly from an experimental perspective only. These directly forced waves, which, in principle, constitute a new initial condition for the parametric instability, appear at $\epsilon^2$ of our asymptotic expansion and they are coupled at order $\epsilon^3$ with the parametric waves, thus influencing not only the wave amplitude saturation, but also the marginal stability boundaries (through a modification of the slope of transition curves) as well as the solution outside the instability regions. If, indeed, for $\theta_s=90^{\circ}$ no meniscus is present and the sub--harmonic parametric waves bifurcate from the flat surface state, when $\theta_s\ne90^{\circ}$, the instability emerges on the top of a still stable, but stationary oscillating free surface, i.e. edge or meniscus waves. This translates in a so-called imperfect bifurcation diagram, which shows a tailing effect owing to meniscus waves, whose amplitude is proportional to the external acceleration amplitude. On this regard, different considerations were made by analogy with previous experimental observations \citep{batson2013faraday}, although for different fluid system and contact line condition. The major influence of contact angle effects on the wave amplitude response was found to occur for axisymmetric sub--harmonic waves. Intuitively, this was explained by considering that harmonic meniscus waves, being directly forced by the spatially-uniform forcing, are axisymmetric by construction, therefore axisymmetric parametric waves, although of different nature, are more likely to be destabilized by edge waves, as they share the same spatial symmetries. This effect is expected to be dominant for harmonic axisymmetric parametric waves, as proved experimentally by \cite{batson2013faraday}. Furthermore, the existence of a harmonic meniscus wave state, from which the parametric waves bifurcate (rather than the flat interface rest state), has been observed in some cases to induce a change of sign of the direction in the bifurcation diagram as the contact angle is varied. Specifically, in some cases the present analysis predicts the existence of a static contact angle for which the bifurcation is always supercritical no matter what the combination of external forcing amplitude and frequency be, thus leading to a suppression of the sub--criticality of the system. This does not seem to have been reported in the literature and it could be tentatively checked in future lab-scale and controlled Faraday experiments.\\
\indent Lastly, with the purpose of validation only, the single-mode WNL model, in the specific case of an axisymmetric dynamics, was compared with fully nonlinear axisymmetric direct numerical simulations (where non-axisymmetric parametric instabilities are artificially filtered out due to axisimmetry), which revealed a good agreement, proving (at least partially) the correctness of the WNL prediction when contact angle effects were introduced.\\
\indent To conclude, we add that the numerical tools developed in this work could enable us to explore different geometries, to revisit previous experiments with different contact line boundary conditions, e.g. the more involved sliding contact line condition (which would require the regularization of the well-known contact line stress-singularity, most likely via phenomenological slip length models \citep{ting1995boundary,miles1990capillary}), to introduce in the latter dynamical contact angle effects \citep{Viola2018a,Viola2018b} and to explore different fluid systems of interest, e.g. multilayer configurations as those investigated by \cite{batson2013faraday}. Moreover, with the aim at quantifying contact angle effects on the Faraday thresholds, the \textit{ad hoc} asymptotic scaling for sub--harmonic parametric resonances defined in the present weakly nonlinear analysis could be modified so to tackle other type of resonances, such as harmonic and super--harmonic parametric waves, combination resonances (see \cite{Kidambi2013}), internal resonances (see Appendix~\ref{sec:appB}) as well as secondary--drift instabilities triggered by pure viscous modes (see Appendix~\ref{sec:appC}). Some of these directions are being pursued and will be reported elsewhere.\\



\appendix

\section{Reconstruction of the full sub--harmonic spectrum discussed by \cite{shao2021surface}}\label{sec:appB}

\begin{figure}
    \centering
    \includegraphics[width=1\textwidth]{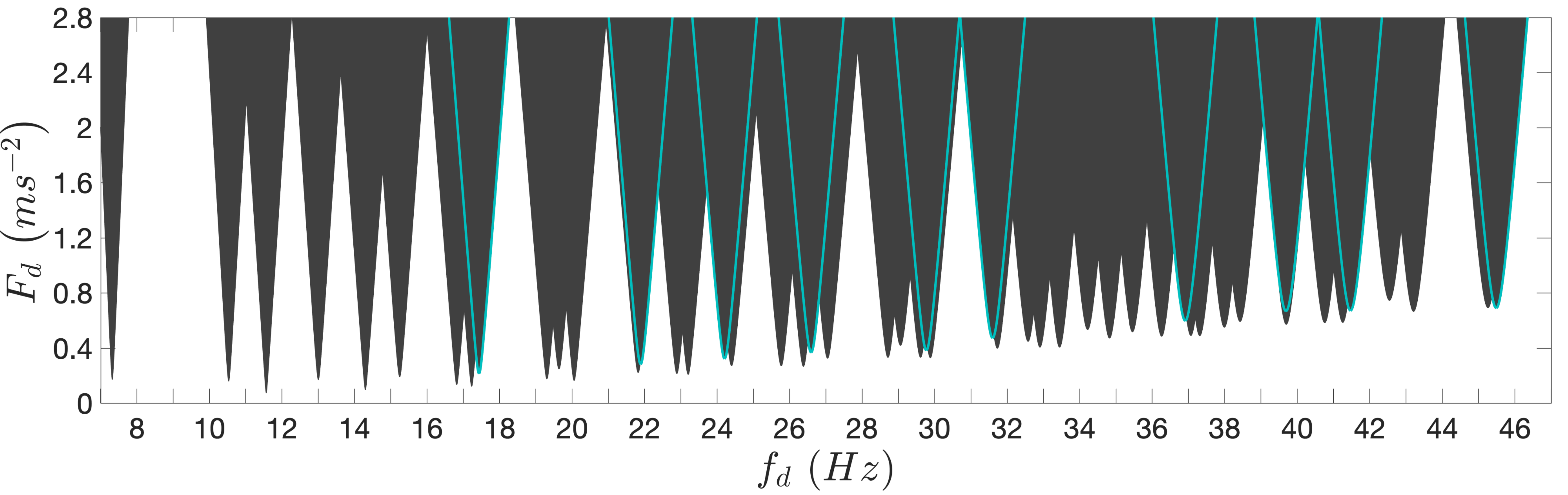}
    \caption{Reconstruction of the full sub--harmonic spectrum in the frequency range considered by \cite{shao2021surface} (S21) in ideal brimful conditions, i.e. $\theta_s=90^{\circ}$ (static--meniscus--free configuration). The boundaries of those Faraday tongues which are less distinguishable, as they almost completely lay within other tongues, are highlighted with aquamarine solid lines and they correspond (ordered in ascendent frequencies) to modes $\left(m,n\right)=\left(5,1\right)$, $\left(7,1\right)$, $\left(8,1\right)$, $\left(9,1\right)$, $\left(7,2\right)$, $\left(11,1\right)$, $\left(13,1\right)$, $\left(14,1\right)$, $\left(11,2\right)$ and $\left(0,6\right)$. }
    \label{fig:full_subharm_spectrum}
\end{figure}

\begin{figure}
    \centering
    \includegraphics[width=0.54\textwidth]{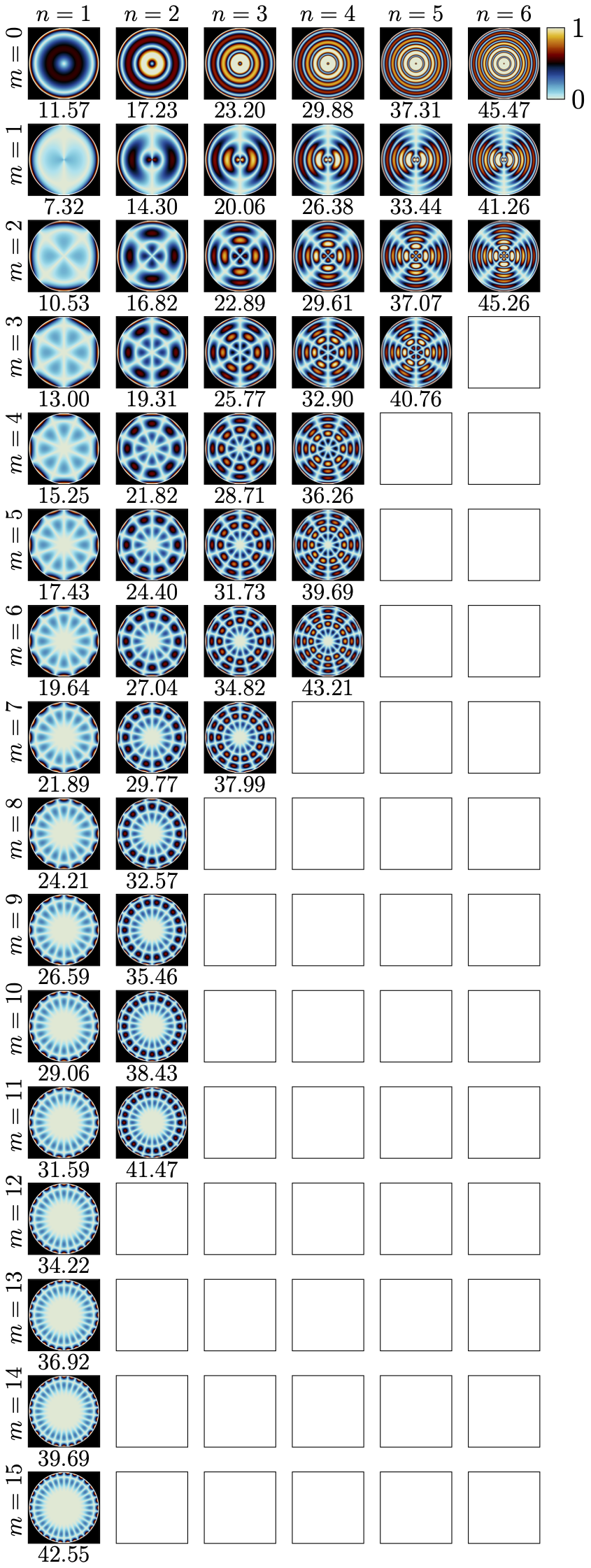}
    \caption{Absolute value of the normalized eigensurface slope for the 50 modes associated with the Faraday tongues of figure~\ref{fig:full_subharm_spectrum}, which reproduce experimental observations by \cite{shao2021surface}. The resonance frequency, $f_d\approx2\omega/2\pi$, at the bottom of each subplot, is in Hz. }
    \label{fig:full_subharm_modes}
\end{figure}

Despite \cite{shao2021surface} (S21) could observe up to 50 different modes in their experiments (whose characteristic frequency ranges are indicated in their figure~5 and are compared with their numerical prediction in their table~1), they reported complete measurements for only 10 sub--harmonic Faraday tongues, corresponding to those discussed in figure~\ref{fig:tongues_teta90}(a) of the present paper. We can therefore use equation~\eqref{eq:WNLeps3_stab_bound} to tentatively reconstruct the entire spectrum in the frequency range $f_d\in\left[7,47\right]$ Hz in order to check whether the present numerical predictions are consistent with S21. This is done in figures~\ref{fig:full_subharm_spectrum} and~\ref{fig:full_subharm_modes}. As aforementioned in \S\ref{subsubsec:Sec5subsec6subsubsec2}, little discrepancies in frequency, when compared with experiments, are observed. Nevertheless, figure~\ref{fig:full_subharm_spectrum} shows that 

\clearpage
\noindent the numerical solution through Eq.~\eqref{eq:WNLeps3_stab_bound} prescribes a well--defined mode quantization in the whole frequency range considered (with increasing onset acceleration for $f_d>25$ Hz), in agreement with S21. According to the present calculation, only the sub--harmonic tongue associated with mode $\left(14,1\right)$ (hardly detectable) completely lies within that of mode $\left(5,4\right)$, although the corresponding instability could be experimentally observed. The associated resonance frequencies and eigensurfaces are displayed in figure~\ref{fig:full_subharm_modes}, which aims to reproduce figure~5 of S21.

\section{Further comments on the imperfect bifurcation diagram and possible internal resonances}\label{sec:appB}

\begin{figure}
    \centering
    \includegraphics[width=1.\textwidth]{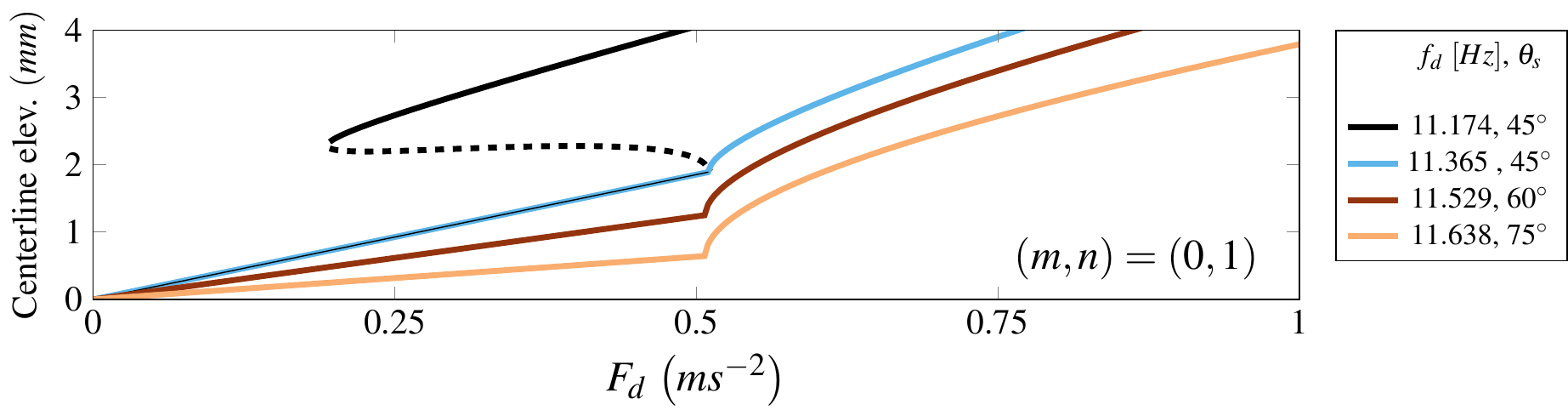}
    \caption{Bifurcation diagrams associated with $\left(m,n\right)=\left(0,1\right)$ (see also figure~\ref{fig:tongues_teta90}(a) and (b)) and for different static contact angles $\theta_s$. The total dimensional centerline elevation (axisymmetric dynamic) is reconstructed by summing the various order solutions, i.e. $\eta=\eta_0+\eta_1+\eta_2$ and it is plotted versus the external forcing acceleration for a fixed excitation angular frequency. The dashed black line for $\theta_s=45^{\circ}$ and $f_d=11.174\,Hz$ denotes the associated unstable branch. Only the super-critical branches are shown for $\theta_s=60^{\circ}$ and $75^{\circ}$. The tailing effect (imperfect bifurcation diagram) produced by presence of harmonic meniscus waves (the amplitude of meniscus waves grows linearly with $F_d$, independently of the parameter combination $\left(F_d,\Omega_d\right)$), is well visible below the Faraday threshold.}
    \label{fig:extra_tailing}
\end{figure}

\begin{figure}
    \centering
    \includegraphics[width=1.\textwidth]{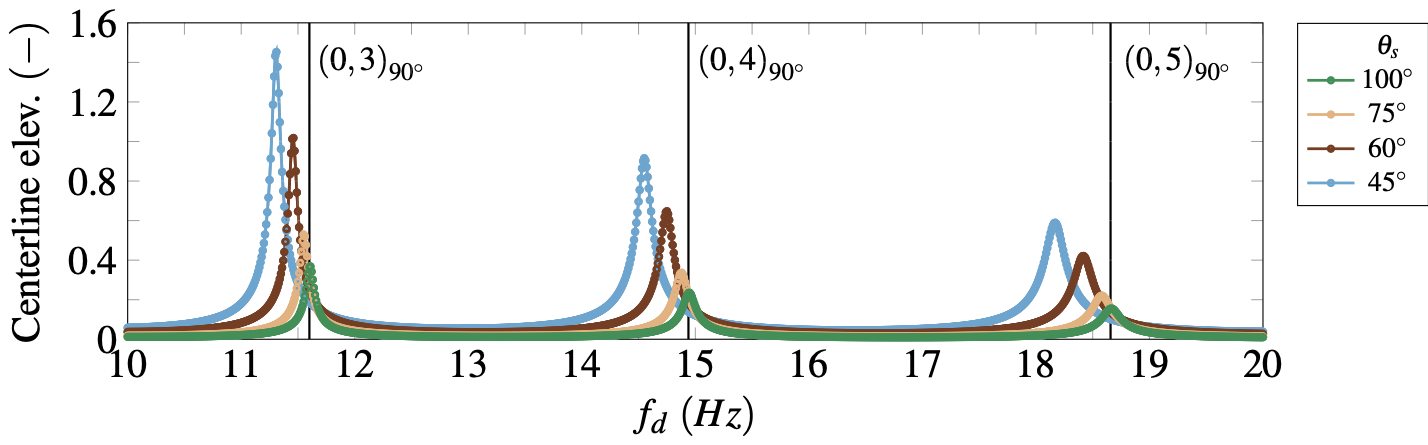}
    \caption{Second order linear response to the external forcing, i.e. meniscus waves, computed by solving~\eqref{eq:WNLeps2_lin_forc_sys} for $\hat{\boldsymbol{\mathcal{F}}}_2^{ij}=\hat{\boldsymbol{\mathcal{F}}}_2^{\hat{F}}$ with a generic varying forcing angular frequency $\Omega_d$ in the frequency window $f_d\in\left[10,20\right]\,\text{Hz}$ (see in combination with figure~\ref{fig:tongues_teta90}). The response is shown in terms of non-dimensional centerline elevation, as the meniscus wave dynamics is axisymmetric. The bifurcation diagram associated with the sub--harmonic axisymmetric wave $\left(m,n\right)=\left(0,1\right)$ displayed in figure~\ref{fig:extra_tailing} is computed at a driving frequency $f_d$ close to that of the axisymmetric natural mode $\left(0,3\right)$. Here, the vertical black solid lines denote the value of the natural frequencies associated with modes $\left(0,3\right)$, $\left(0,4\right)$ and $\left(0,5\right)$ computed for a static contact angle $\theta_s=90^{\circ}$.}
    \label{fig:lin_men_wave_resp}
\end{figure}

Although not investigated here, the asymptotic model formalized in this work allows us to enlighten some important points regarding harmonic axisymmetric parametric waves \citep{batson2013faraday}. Axisymmetric harmonic meniscus waves, which in our asymptotic scaling for sub--harmonic parametric resonances do not resonate at order $\epsilon^2$, even for $\left(m,n\right)=\left(0,n\right)$ modes, will immediately resonate at the $\epsilon^2$--order if harmonic axisymmetric parametric waves are considered, $\Omega_d\approx\omega_{0n}$ ($\left(m,n\right)=\left(0,n\right)$). In order to be tackled, this peculiar case would require the imposition of a second order solvability condition, which generates an additive forcing term, $\sim F$, in contrast to a typical multiplicative forcing term, e.g. $\sim FB^*$ for sub--harmonic resonances as those examined the present paper, typical of parametric instabilities. This suggests the appearance of a dominant (of order $\epsilon^2$ instead of $\epsilon^3$ or higher) Duffing--like term in the corresponding final amplitude equation. In other words, the system undergoes a combined (dominant) direct and parametric (higher order) resonance owing to resonant meniscus waves. Although further detailed investigations of this specific situation must be pursued, this effect is expected to enhance both the tailing effect and the harmonic parametric instability and, therefore, to lower the detectable threshold of these harmonic standing capillary--gravity waves, as experimental observations by \cite{batson2013faraday} confirm.\\
\indent Furthermore, the excitation of harmonic meniscus waves strongly enhances the phenomenon of internal resonances \citep{Miles90,Miles84,Nayfeh1987,Meliga2012}. As an illustrative example on this regard, in figure~\ref{fig:extra_tailing} the bifurcation diagram associated with the sub--harmonic axisymmetric parametric waves $\left(m,n\right)=\left(0,1\right)$ (see also figure~\ref{fig:tongues_teta90}(a)) is shown. In the configuration of figure~\ref{fig:tongues_teta90}(a) and figure~\ref{fig:extra_tailing}, the sub--harmonic for $\left(0,1\right)$ occurs at a driving frequency $f_d\approx11.4\,\text{Hz}\approx2\omega_{01}$ (the exact value depends on the value of $\theta_s$). As shown in figure~\ref{fig:lin_men_wave_resp}, the angular frequency $2\omega_{01}^{\theta_s}$ happens to be close to the natural frequency of an axisymmetric mode, i.e. $\omega_{03}$. In this case an internal resonance is expected to take place. In a meniscus--free configuration, the internal resonance would be produced by the interaction of a sub--harmonic and harmonic parametric waves \citep{Miles90,Miles84,Nayfeh1987,Meliga2012}. Nevertheless, unlike the classic case, presence of meniscus and, therefore, of meniscus waves, strongly favors such a mechanism, i.e. the harmonic resonance does not correspond to a pure parametric wave, but rather to a meniscus--driven wave (additive forcing). Figures~\ref{fig:extra_tailing} and~\ref{fig:lin_men_wave_resp} enlighten how the vicinity to an axisymmetric natural mode dramatically enhances the tailing effect. However, in spite of the fact that results proposed in figure~\ref{fig:extra_tailing} are qualitatively meaningful, it should be noted that the present WNL model assumes meniscus waves to be smaller than the leading parametric wave, which is no more the case in such conditions. Hence, despite viscosity helps by progressively damping the meniscus wave response for increasing frequency, the present analysis should not be considered quantitatively reliable for driving frequency close to those natural frequencies associated with axisymmetric modes, for which the order of magnitude of the second order meniscus wave response is expected to become comparable to that of the leading order solution, thus breaking the asymptotic expansion. In these case a more subtle formulation of the leading order problem, in order to avoid harmonic resonances and therefore secular terms at second order, based on a two mode expansion, should be retained \citep{Meliga2009,Meliga2012,bongarzone2021impinging} (beyond the scope of this work though). For these reasons, the DNS analysis reported in \S\ref{sec:Sec6} has been applied to mode $\left(0,2\right)$ rather than $\left(0,1\right)$ so to check more precisely the validity of the WNL analysis.


\section{Discussion on secondary-drift instability due to pure viscous modes}\label{sec:appC}

For slightly viscous cases as those examined in this paper, \cite{martel1997damping} have pointed out that pure viscous modes (with zero oscillation frequency $\omega=0$) could have smaller damping rates than the capillary--gravity ones and may be therefore important for an accurate prediction of the Faraday instability. Although in the present analysis such modes have been ignored, the comparison with experiments and axisymmetric direct numerical simulations shows generally a fairly good agreement, suggesting that their influence is not significant in the cases that we have analyzed. Nevertheless, some crucial aspects are worth to be discussed.\\
\indent The second order response discussed in \S\ref{sec:Sec5} includes several contributions, among which the time- and azimuthal-averaged flows, i.e. steady and steady-axisymmetric mean flows, associated with the two perfectly balanced counter-rotating waves. Within the present WNL model, these steady forcing terms have been considered non-resonant and the corresponding responses have been therefore computed straightforwardly. However, from a mathematical perspective, one must notice that existence of viscous modes, depending on the associated damping coefficient, may induce large second order mean flow responses when system~\eqref{eq:WNLeps2_lin_forc_sys} is inverted for the couple $\left(m^{ij},\omega^{ij}\right)=\left(m,0\right)$ or $\left(0,0\right)$, i.e. the system could be nearly singular if the damping coefficients of some of the first viscous modes are sufficiently small and therefore close to neutrality. It is fundamental to note that for $\left(m^{ij},\omega^{ij}\right)=\left(0,0\right)$ the linearized governing equations decouple into two sets of equations, one governing the vertical-plane dynamics, $r$-$z$, and one governing the azimuthal velocity component only. Therefore, the steady axisymmetric mean flow responses can be interpreted as the sum of two contributions, i.e. an in-plane or toroidal mean flow in the $r$-$z$ plane (analogous to that recently studied experimentally by \cite{perinet2017streaming} with regard to a three-dimensional rectangular container) and a pure azimuthal or drift mean flow component. When a standing wave solution is considered, the second order toroidal mean flows produced by the two counter rotating waves sum up, whereas the two opposite azimuthal mean flows cancel out, leading to a net zero drift. However, through an amplitude equation model capturing the infinite number of viscous modes at leading order (together with the classic standing wave form), it has been shown \citep{fauve1991drift,knobloch2002coupled,martel2000dynamics,vega2001nearly} that in some conditions the amplitude of one of the two azimuthal mean flow component may exceed that of the opposite mean flow, hence inducing, particularly at larger standing wave amplitudes, a secondary instability via space-reflection symmetry breaking.\\
\indent An exception is made by standing waves produced by the combination of two axisymmetric, $m=0$, capillary--gravity waves, as that analyzed in \S\ref{sec:Sec6} for $m=0$ and $n=2$. In this specific case, the resulting second order azimuthal mean flow is zero by construction precluding the existence of such a secondary instability and the overall mean flow acts toroidally (in the $r$-$z$ plane) exclusively. In other words, although the influence of viscous modes should not be neglected \text{a priori}, these arguments explain how standing wave dynamics produced by axisymmetric capillary--gravity waves, as the one investigated through DNS in \S\ref{sec:Sec6} of the present manuscript, do not suffer from secondary drift instability.


\subsubsection*{\textbf{\textup{Supplementary Material}}}
Wolfram Mathematica codes developed in this work for the automatized linearization process and specifically for extraction of the WNL second order forcing terms and third order resonating terms, discussed in \S\ref{sec:Sec5}, are available to the readers as a supplementary material.


\subsubsection*{\textbf{\textup{Acknowledgements}}}
\indent The authors wish to thank S. Gomé for his experimental observations that have motivated this work as well as Dr. L. Siconolfi for his support.

\subsubsection*{\textbf{\textup{Funding}}}
\indent We acknowledge the Swiss National Science Foundation under grant 200021\_178971.\\

\subsubsection*{\textbf{\textup{Declaration of Interests}}}
\indent The authors report no conflict of interest.

\bibliographystyle{jfm}
\bibliography{Bibliography}

\end{document}